\documentclass[english,aps,reprint,nofootinbib,onecolumn]{revtex4-2}
\usepackage{lmodern}

\usepackage[T1]{fontenc}
\usepackage[latin9]{inputenc}
\usepackage{xcolor}
\usepackage{calc}
\usepackage{dsfont}
\usepackage{amsmath}
\usepackage{graphicx}
\PassOptionsToPackage{normalem}{ulem}
\usepackage{ulem}

\makeatletter

\providecolor{lyxadded}{rgb}{0,0,1}
\providecolor{lyxdeleted}{rgb}{1,0,0}
\DeclareRobustCommand{\mklyxadded}[1]{\textcolor{lyxadded}\bgroup#1\egroup}
\DeclareRobustCommand{\mklyxdeleted}[1]{\textcolor{lyxdeleted}\bgroup\mklyxsout{#1}\egroup}
\DeclareRobustCommand{\mklyxsout}[1]{\ifx\\#1\else\sout{#1}\fi}

\usepackage{dsfont}

\makeatother

\usepackage{babel}
\begin{document}
\title{Fixed points of Boolean networks with sparse connections}
\author{Stav Marcus, Ari M. Turner, Guy Bunin}
\address{Technion-Israel Institute of Technology, Haifa, Israel}
\author{Bernard Derrida}
\address{Collège de France, 11 place Marcelin Berthelot, 75005 Paris, France~\\
 Laboratoire de Physique de l\textquoteright ENS, UMR 8023, CNRS -
ENS Paris - PSL University - SU - Université Paris Cité, Paris, France}
\begin{abstract}
We study fixed points of cellular automata with $N$ sites on random
sparse graphs. In the large $N$ limit such models are known to exhibit
phase transitions, from a ``frozen'' phase, where at most a finite
number of sites fluctuate at long times, to a ``fluctuating'' phase
where a finite fraction of sites fluctuate. We consider several models,
calculating the first and second moments of the number of fixed points,
and find that these moments remain finite in the large $N$ limit,
except at the transitions where they become singular. The singularities
can take several forms, including divergence of the mean or variance
of the number of fixed points, on one or both sides of the transition.
The type of singularity is related to properties of the mean field
dynamics or dynamics of the distance between copies of the system.
In configuration space, we find that fixed points are organized into
clusters, each consisting of sets of fixed points that agree with
one another except for on a finite number of sites. In the frozen
phase there is only one cluster, while in the fluctuating phase there
may be multiple clusters. If there are multiple clusters, the distance
between fixed points in different clusters is extensive. We show that
the differences within the clusters correspond to local changes near
short cycles in the directed graph of connections whose influence
is eventually limited. In the frozen phase, we calculate the full
distribution of the number of fixed points.
\end{abstract}
\maketitle

\section{Introduction}

Fixed points are objects of prime interest in dynamical systems. One
would like to know how many there are, how they are organized in phase
space, their stability, their basins of attraction and more. Here
we study properties of fixed points (FPs) in cellular automata. Cellular
automata are useful models in many fields, for example as models of
disordered systems, neural networks, ecological networks, and gene
expression networks \citep{mezard_spin_1987,kauffman_metabolic_1969,kauffman1969homeostasis,hopfield_neural_1982,fisher_transition_2014}.

We consider a set of sites $i=1,...,N$, where the value at site $i$
at time $t$ is a binary variable, $x_{i}(t)\in\{0,1\}$. We will
also refer to a site $i$ as being ``on'' or ``off'' for $x_{i}=1,x_{i}=0$
respectively. Time is taken to be discrete, and at each time step
the sites are updated by the rule $x_{i}(t+1)=f_{i}\left(x_{j_{1}}(t),x_{j_{2}}(t),...\right)$,
where $f_{i}$ is a function of the values at a finite number of input
sites $x_{j_{n}}(t)$. Both the set of input sites to $i$ and the
function $f_{i}$ are fixed in time. In the networks considered in
the present paper, the connections are chosen randomly and independently,
and the functions are also chosen randomly from some ensemble, depending
on the choice of model. This kind of system can be represented by
a directed graph, with the sites represented as vertices and a link
$j\rightarrow i$ if $x_{j}$ is an input to $f_{i}$. We will be
interested in the limit where the system has many sites, $N\gg1$,
but the graph of interactions remains sparse, so that the distribution
of the number of input sites is independent of $N$. Due to sparseness,
most interactions are ``one-way'': if $f_{i}$ depends on some $x_{j}$,
then with high probability $f_{j}$ does not depend on $x_{i}$.

Many cellular automata exhibit dynamical phase transitions, from a
``frozen'' phase, where for sparse random and independent connections,
the dynamics of all but possibly a finite number (i.e., independent
of $N$) of variables reach a fixed value, to a fluctuating or chaotic
phase, where a finite fraction of the sites change their values indefinitely
\citep{derrida_dynamical_1987,bastolla_relevant_1998,lynch_threshold_1995,Derrida_Distance_Evolution_1986,Derrida_Local_Magnetisations_1987}.
The Kauffman model, for example, which models the interactions of
genes which can regulate one another's activation state \citep{kauffman1969homeostasis,kauffman1969metabolic},
has such a transition to a phase with chaotic dynamics. For a given
phase, frozen or fluctuating, the system will reach the corresponding
dynamical behavior (fixed in time or fluctuating, respectively) almost
surely for any randomly chosen initial conditions. However, the system
may have FPs even in the fluctuating phase that can only be reached
starting from special initial conditions.

We will be interested in understanding the fixed points in both phases
of this phase diagram, and in particular, the probability distribution
of the number of fixed points. Besides considering how many fixed
points there are we will consider how they are organized in phase
space. We will see that the answers are intimately related to the
dynamics in the two phases. The idea that this might happen is inspired
by considering the Kauffman model, where the number of limit cycles
(also known as attractors) diverges in a special way as a function
of $N$ at the critical point \citep{Bastolla1997,Socolar2003,samuelsson_superpolynomial_2003,klemm_stable_2005,greil_dynamics_2005,krawitz_basin_2007,Kaufman2005,mihaljev2006,fink_number_2023}.
This suggests that as the critical point is approached, there can
be a singularity in the distribution of the number of \emph{fixed
points}, since they are just cycles of period 1. We derive formulae
relating the statistics of the number of fixed points to dynamics
in a precise way (generalizing a formula in Ref. \citep{samuelsson_superpolynomial_2003})
and we use it to calculate various properties of the number of fixed
points.\textcolor{orange}{{} }We will also look beyond the Kauffman
model in order to see whether there are different types of behavior
in different models. For example, is the behavior universal for all
phase transitions from frozen to chaotic phases? The models we consider
all consist of networks with a specific type of random one-way connections,
similar to Erdos-Renyi directed graphs. An interesting property of
this family of models is that the number of FPs is of order 1, except
at transitions. Ref. \citep{fried_alternative_2017} noticed this
for a model (which we call the inhibitory model): when this model
is considered on an an Erdos-Renyi graph with one-way connections,
the average number of FPs is independent of the system size, in contrast
to the case of a graph where the probability of having an edge from
variable $i$ to $j$ is correlated with having an edge from $j$
to $i$.

In a frozen phase, we will be able to understand the fixed points
by reducing the problem to finding FPs for small disconnected parts
of the interaction graph, for which it is easy to determine the number
of FPs by going through all combinations of values. This method leads
to a determination of the full probability distribution for the number
of fixed points. In a fluctuating phase, this method does not work;
in this phase it is even difficult to count the number of fixed points
on a computer, since the only way to find them seems to be to test
exponentially many (in $N$) combinations of the variables. Neverthless,
we will be able to calculate \emph{moments} of the number of fixed
points. The method for this is a discrete-variable counterpart of
the Kac-Rice technique that applies to systems of interacting \emph{continuous}
variables. Such a method has been applied previously to the Kauffman
model \citep{samuelsson_superpolynomial_2003} and in the context
of a specific model for ecology \citep{fried_alternative_2017}. This
method gives formulae that determine the expected number of fixed
points in terms of a function that describes the evolution of the
mean field density for generic initial conditions, as well as generalizations
that relate higher moments to the time dependence of correlations
between several configurations.

These questions can be generalized to other transitions besides the
one from a frozen to a chaotic phase. Besides the Kauffman model,
we study three other models (which are of the ``regulatory type''
studied by \citep{greil_dynamics_2005,aracena2008}) to explore various
possible types of behaviors, in particular around the dynamical transition.
These are described in the box below. We find that there can be different
behaviors in the number of fixed points for the different models,
which can be classified into universality classes, and that the behavior
is connected to the way the dynamics changes at the transition. In
particular, in some cases, which include the Kauffman and the inhibitory
model (see Box), the \emph{mean} number of fixed points remains finite
at the dynamical transition and is even analytic there, but the \emph{probability}
for a fixed point to exist goes to zero, and the \emph{second moment}
diverges! How these properties are consistent with each other can
be understood from the full distribution of the number of fixed points;
on the other hand, the moment formulas give a natural explanation
for why the singularity is not seen until the second moment in terms
of the relationship between moments and the dynamics.

\medskip{}

\noindent\fcolorbox{black}{white}{\begin{minipage}[t]{1\columnwidth - 2\fboxsep - 2\fboxrule}%
The variables $x_{i}$ in all models have two values, denoted by $\{0,1\}$.
All models have one parameter, $C$, the mean number of incoming edges
to each site, with edges from a vertex to itself allowed. The system
size (number of sites) is $N$. The number of fixed points (FPs),
$\Omega$, is a random variable that depends on the realization of
the disorder. The models considered are:
\begin{itemize}
\item Kauffman model: the function determining $x_{i}(t+1)$ chosen at random,
uniformly from functions with appropriate number of inputs.
\item Inhibitory model: $x_{i}(t+1)=1$ iff for all incoming edges (from
$j$ to $i$) $x_{j}(t)=0$.
\item Excitatory model: $x_{i}(t+1)=1$ iff at least one incoming edge (from
$j$ to $i$) has $x_{j}(t)=1$.
\item Double excitatory model: $x_{i}(t+1)=1$ iff at least TWO of the incoming
edges have $x_{j}(t)=1$.
\end{itemize}
\end{minipage}}

This paper is organized as follows. In section 2, we consider the
Kauffman model, for which we compute the first and second moments
of the number of fixed points. We observe that in the frozen phase,
all the fixed points are concentrated in a single cluster, whereas
in the fluctuating (i.e. chaotic phase), they are grouped into several
clusters at a distance from each other related to the fixed points
of the evolution of the overlap fraction. At the transition between
the frozen and the fluctuating phase, the second moment of the number
of fixed points diverges, growing like $N^{\frac{1}{3}}$ with the
number $N$ of sites. In the frozen phase, the analysis of short cycles
in the graph allows one to give the physical origin of these results
and to obtain the full distribution of the number of fixed points.
In section 3, we first show that the approach developed in the previous
section to calculate the moments of the number of fixed points can
be generalized to other sparse networks of automata. In particular
we derive a relation Eq. (\ref{eq:saddlepoint-1}) between the first
moment and the evolution $\psi\rightarrow q(\psi)$ of the magnetization
(see Appendix A for the derivation), and we discuss the models just
listed. We find various differences in the behavior of the moments,
corresponding to the difference in their dynamical transitions. In
particular, the double excitatory model has a first order transition
in its dynamics which causes the first moment to diverge in a different
way as a function of $C-C_{c}$ and $N$ than in models with second
order transitions.

\section{Some of the main phenomena in a single model}

\subsection{Kauffman model: definition and dynamical phases\label{subsec:Kauffman-model:-definition}}

We first introduce some of the main phenomena through the example
of one model, the Kauffman model \citep{kauffman_metabolic_1969}.
In the variant of the Kauffman model that we use, one first constructs
a random Erd\H{o}s\textendash Rényi directed graph with a mean of
$C$ edges as input connections to each sites. This is done by adding
each edge $j\rightarrow i$ to the graph with a probability $C/N$,
including allowing self loops where $j=i$. The inclusion of loops\footnote{Below ``loops'' refer to edges that connect a vertex to itself,
and cycles to refer to closed paths of any length.} makes results more compact, and does not affect the singular behavior
at and near the transitions (see Appendix \ref{sec:InhibitoryAppendix}
for the effects of loops in one model). For large $N$, which will
be assumed throughout, the number $k$ of inputs to a site is therefore
approximately a Poisson random variable with probabilities $P_{k}=e^{-C}C^{k}/k!$,
independently for different sites $i$. The function $f_{i}$ that
determines $x_{i}(t+1)$ is chosen at random, uniformly from the $2^{2^{k}}$
boolean functions with $k$ inputs (with the convention that if there
are no inputs to $i$, $f_{i}$ is a constant, and $f_{i}=0$ or $f_{i}=1$
with equal probability). The connections define a directed graph,
with an edge $j\to i$ if $x_{j}$ is an input to $f_{i}$.

Before discussing FPs, we briefly review some results regarding the
dynamical behavior of the model. The model features two dynamical
phases \citep{derrida_random_1986,lynch_threshold_1995}, depending
on $C$. To see this, we would like to compare the evolution of two
configurations of the system initialized with different initial conditions:
will they tend to diverge from each other or not?

For this purpose, consider two configurations at time $t$, $x(t),y(t)\in\{0,1\}^{N}$
(where $x=\left\{ x_{i}\right\} $ is a binary vector of the values
at all sites). Let the overlap fraction $\phi_{t}$ be the fraction
of sites for which $x_{i}(t)=y_{i}(t)$. We calculate $\phi_{t+1}$
in the limit of large $N$ \citep{derrida_random_1986}. Consider
the sites $i$, which can be grouped into two types. The first type
are those such that all inputs are equal in $x$ and $y$, i.e. all
$j\to i$ satisfy $x_{j}(t)=y_{j}(t)$, which implies that $x_{i}(t+1)=y_{i}(t+1)$.
For a site with $k$ input sites, the probability that all inputs
are equal is $\left(\phi_{t}\right)^{k}$, and averaging over $k$,
the fraction of sites of this type is
\[
\sum_{k=0}^{\infty}P_{k}\phi_{t}^{k}=e^{C(\phi_{t}-1)}\ .
\]
The remaining sites are those such that $x_{j}(t)\ne y_{j}(t)$ for
at least one of the input sites $j\to i$. In this case, $x_{i}(t+1)$
and $y_{i}(t+1)$ are values of $f_{i}$ for two distinct argument-sets,
and since $f_{i}$ is uniformly sampled from all possible boolean
functions, the probability that $x_{i}(t+1)=y_{i}(t+1)$ is $1/2$.
Taking these two possibilities into account, we find that
\begin{equation}
\phi_{t+1}=q(\phi_{t})\label{eq:q_definition}
\end{equation}
with 
\begin{equation}
q(\phi_{t})=e^{C(\phi_{t}-1)}+\frac{1}{2}\left[1-e^{C(\phi_{t}-1)}\right]=\frac{1}{2}\left[1+e^{C(\phi_{t}-1)}\right]\ .\label{eq:q_Kauffman}
\end{equation}
Analyzing this recursion relation, we find a transition in the behavior
at $C_{\text{crit}}=2$. We see that for $C<2$ it admits only one
fixed point, at $\phi=1$, which is attractive. Therefore, any pair
of initial conditions that are different from each other eventually
reach an overlap fraction of 1; that is they become equal to one another,
apart from a sub-extensive number of sites. Hence all such initial
conditions converge to the same final state, up to a subextensive
number of sites. This is known as the frozen phase. For $C>2$, the
fixed point at $\phi=1$ becomes repulsive, and there is another attractive
fixed point at some $0<\phi^{*}<1$, given by 
\begin{equation}
\phi^{*}=\frac{1}{2}-\frac{1}{C}W\left(-\frac{C}{2}e^{-C/2}\right)\ .\label{eq:Kaufmann_phi_star_expression}
\end{equation}
Here $W(\zeta)$ is the Lambert W-function, defined as the solution
to the equation $We^{W}=\zeta$. (When $\zeta<0$ the larger of the
two solutions is taken. Several useful facts on this function are
collected in Appendix \ref{sec:Lambert_W}.) Thus, similar initial
conditions ($\phi$ close to $1$) diverge, indicating that this is
a fluctuating phase.
\begin{figure}
\begin{centering}
\includegraphics[width=0.7\textwidth]{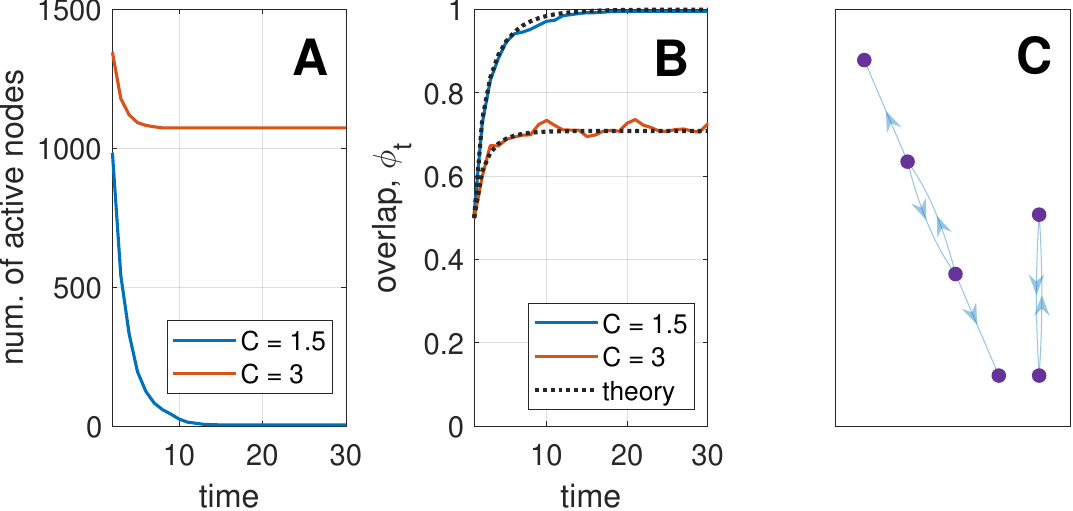}
\par\end{centering}
\caption{\textbf{\label{fig:Dynamics}Dynamics of the Kauffman model above
and below }$C_{\text{crit}}=2$\textbf{.} Simulations with $N=1500$.
(A) The number of active nodes. For $C<C_{\text{crit}}$ (here $C=1.5$),
this number may go to zero if a fixed point is reached, or in this
simulation, to a small positive number. For $C>C_{\text{crit}}$,
it stabilizes on a finite fraction of the nodes. (B) The overlap goes
close to $\phi_{t}=1$ for $C<C_{\text{crit}}$, and for $C>C_{\text{crit}}$
fluctuates around some value (at $N\to\infty$ this is $\phi^{*}$,
see Eq. (\ref{eq:Kaufmann_phi_star_expression})). The dotted lines
are the solutions of the $\phi_{t}$ dynamics, Eqs. (\ref{eq:q_definition},\ref{eq:q_Kauffman}).
(C) An example of the structure of the subgraph comprised of the active
nodes for $C<C_{\text{crit}}$. It consists of several short cycles
(here, two cycles of length two), and nodes that are downstream from
them.}
\end{figure}

To see what these results imply for the dynamics of a system at finite
$N$, we perform simulations at two values of $C$, below and above
$C_{\text{crit}}=2$. See Fig. \ref{fig:Dynamics}, where $N=1500$,
and $C=1.5,3$. Shown are the number of active variables as a function
of time, where an active variable is one that continues to change
indefinitely (in the simulation, until some large time), and the overlap
$\phi_{t}$ between runs from independent random initial conditions.

For $C<C_{\text{crit}}$, the number of active sites decreases until
either reaching a fixed point, or stabilizing on some small number
of active sites, see Fig. \ref{fig:Dynamics}(A). Plotting the directed
graph reduced to these active sites, shows that they include a cycle
or several cycles, and sites that are downstream from the cycle, see
Fig. \ref{fig:Dynamics}(C). The overlap increases to include all
sites except some of these active sites, Fig \ref{fig:Dynamics}(C).
In this phase, we show below that at large $N$, all FPs (if they
exist) are configurations similar to these long-time fluctuating configurations,
in that they have the same values on all but a finite number of nodes.

When $C>C_{\text{crit}}$, the number of active sites stabilizes on
a finite fraction of $N$, and the overlap fluctuates around some
value between zero and one. Fixed points are not reached in tested
simulation times. Here it is hard to obtain FPs through such dynamics
simulations, however we show that they may exist, and obtain some
of their properties analytically.

\subsection{Moments of the number of fixed points\label{subsec:Kauffman-model_moments}}

We now turn to discuss the number of FPs, and whether it changes in
a way that reflects the dynamical phase transition just found. An
FP is a configuration $x$ such that for all $i$, $x_{i}(t+1)=x_{i}(t)$.
The number of FPs is a random variable, denoted throughout by $\Omega$.

\subsubsection{Calculating the first moment by summing over configurations}

One way to calculate moments of $\Omega$ is based on the following
formula for $\Omega$ for a given disorder realization $D$ (i.e.,
a choice of the connectivity of the graph and the functions $f_{i}$):
\[
\Omega(D)=\sum_{x}\mathds{1}_{\text{\ensuremath{x} is FP of \ensuremath{D}}}\,.
\]
That is, the number of fixed points can be found by summing over all
binary vectors $x=\left\{ x_{i}\right\} $ a term that is equal to
1 if $x$ is a fixed point and 0 otherwise. The mean number of FPs
is thus found by averaging over the different realizations, 
\begin{align}
\left\langle \Omega\right\rangle  & =\sum_{D}P(D)\sum_{x}\mathds{1}_{\text{\ensuremath{x} is FP of \ensuremath{D}}}=\sum_{x}\sum_{D}P(D)\mathds{1}_{\text{\ensuremath{x} is FP of \ensuremath{D}}}\,.\label{eq:mean_omega_as_sum}
\end{align}
 Let us fix the vector $x$ and evaluate the sum over $D$. For any
given $x$, this sum is just the probability, over the choice of connectivity
and of the $f_{i}$'s, that $x$ is a fixed point. In other words,
if $x(t)=x,$ what is the probability that $x(t+1)=x$ as well? Since
the output of a random function $f_{i}$ is equally likely to be $0$
or $1$, the probability that $x_{i}(t+1)=x_{i}(t)$ is $1/2$, and
therefore the probability that $x(t+1)=x(t)$ is $2^{-N}$. So,
\[
\left\langle \Omega\right\rangle =\sum_{x}2^{-N}=1\ .
\]

\subsubsection{Deriving an expression for the second moment by summing over pairs
of configurations}

The second moment is given by
\[
\left\langle \Omega^{2}\right\rangle =\sum_{D}P(D)\left(\sum_{x}\mathds{1}_{\text{\ensuremath{x} is FP of \ensuremath{D}}}\right)^{2}=\sum_{\left\{ x,y\right\} }\sum_{D}P(D)\mathds{1}_{\text{\ensuremath{x} is FP of \ensuremath{D}}}\mathds{1}_{\text{\ensuremath{y} is FP of \ensuremath{D}}}
\]
Here, $x,y$ are two possible configurations, and we need to calculate
the probability over the disorder $D$ that they are both fixed points.

In the following calculation, it will be computationally convenient
to work with a slightly different ensemble of networks, which will
give the same results in the large $N$ limit. For each site $i$,
one first chooses how many inputs $k$ there should be to this site,
and then one chooses the $k$ sites independently from among the vertices
in the graph, allowing multiple edges and loops. If the number of
inputs is chosen according to a Poisson distribution with a mean $C$,
this will be nearly equivalent to the procedure described in \ref{subsec:Kauffman-model:-definition}
when $N$ is large, except for the existence of $O(1)$ multiple edges.
Allowing multiple edges can be checked to have no effect on the numbers
of fixed points in the limit $N\rightarrow\infty$, however.

Let $N_{11}$ be the number of pairs $(x_{i},y_{i})=(1,1)$ at time
$t$, and define similarly $N_{01},N_{10},N_{00}$. Note that $N_{00}+N_{01}+N_{10}+N_{11}=N$.
We now calculate the probability $q_{11}$ that for a randomly chosen
$i$, at the next iteration $x_{i}(t+1)=1$ and $y_{i}(t+1)=1$. Let
$k$ be the number of inputs to site $i$. If $x$ and $y$ are equal
on all inputs to $f_{i}$, $\left(x_{j_{1}},..,x_{j_{k}}\right)=\left(y_{j_{1}},..,y_{j_{k}}\right)$,
then the values at the next iteration are equal, $x_{i}(t+1)=y_{i}(t+1)$,
and are both $1$ with probability $1/2$. Otherwise, if some inputs
are different, then the values $x_{i}(t+1),y_{i}(t+1)$ are values
of $f_{i}$ for distinct arguments, so the probability that they are
both 1 is $1/4$. Hence, if there are $k$ inputs to site $i$, the
probability that $x_{i}(t+1)=y_{i}(t+1)=1$ is
\[
\frac{1}{2}\left(\frac{N_{A}}{N}\right)^{k}+\frac{1}{4}\left[1-\left(\frac{N_{A}}{N}\right)^{k}\right]=\frac{1}{4}\left(1+\phi^{k}\right)\ ,
\]
where $N_{A}\equiv N_{00}+N_{11}$ and $\phi\equiv N_{A}/N$ is the
fraction of sites that are equal\footnote{The preceding formula uses the assumption that multiple edges between
the same vertices are allowed: each randomly chosen input site has
a probability $\frac{N_{A}}{N}$ that $x=y$ at it, so the probability
that this is true for all $k$ input sites is $\left(\frac{N_{A}}{N}\right)^{k}$
because the input sites are chosen independently from one another,
allowing possible repetitions.} on $x,y$. Averaging over the number of inputs $k$,
\begin{equation}
q_{11}=\sum_{k=0}^{\infty}P_{k}\frac{1}{4}\left(1+\phi^{k}\right)=\frac{1}{4}\left[1+e^{C(\phi-1)}\right]\ .\label{eq:q11Kauffman}
\end{equation}
The chances that $x_{i}(t+1),y_{i}(t+1)$ have any other pair of values
can be found in similar fashion: $q_{00}=q_{11}$, and
\begin{equation}
q_{01}=q_{10}=\sum_{k=0}^{\infty}P_{k}\frac{1}{4}\left(1-\phi^{k}\right)=\frac{1}{4}\left[1-e^{C(\phi-1)}\right]\ .\label{eq:q01Kauffman}
\end{equation}

Now for a site $i$ whose value is 11 at time $t$ (i.e., $(x_{i}(t),y_{i}(t))=(1,1)$),
the chance that it will be fixed after one step is the chance that
its value is 11 at time $t+1$, i.e. it is $q_{11}$. As there are
$N_{11}$ such sites, the probability that $x,y$ are both fixed points
is $q_{00}^{N_{00}}q_{01}^{N_{01}}q_{10}^{N_{10}}q_{11}^{N_{11}}$.
Summing over all possible $x,y$ with the appropriate combinatorial
prefactors,
\begin{align}
\left\langle \Omega^{2}\right\rangle  & =\sum_{N_{00},N_{01,}N_{10},N_{11}}\frac{N!}{N_{00}!N_{01}!N_{10}!N_{11}!}q_{00}^{N_{00}}q_{01}^{N_{01}}q_{10}^{N_{10}}q_{11}^{N_{11}}\label{eq:secondmoment-sum}
\end{align}
where the variables summed over are restricted so that $N=N_{00}+N_{11}+N_{01}+N_{10}$.
We simplify this sum by defining $N_{A}=N_{00}+N_{11}$ and summing
over all the $N$'s with a certain value of $N_{A}$. There are two
independent variables since $N_{00}+N_{11}=N_{A}$ and $N_{01}+N_{10}=N-N_{A}.$
Since the $q_{ss'}$ depend only on $N_{A},$ the binomial theorem
can be used to calculate the sums, giving:

\begin{align}
\left\langle \Omega^{2}\right\rangle  & =\sum_{N_{A}=0}^{N}\binom{N}{N_{A}}q(\frac{N_{A}}{N})^{N_{A}}(1-q(\frac{N_{A}}{N}))^{N-N_{A}}\ ,\label{eq: second-moment short sum}
\end{align}
where $q=q_{00}+q_{11}=q(\phi)$ is the same function that appeared
in the \emph{dynamics} of the overlap fraction $\phi_{t}$, Eq. (\ref{eq:q_Kauffman}).

\subsubsection{Evaluating the second moment using the saddle point method}

\begin{figure}
\begin{centering}
\includegraphics[width=0.5\textwidth]{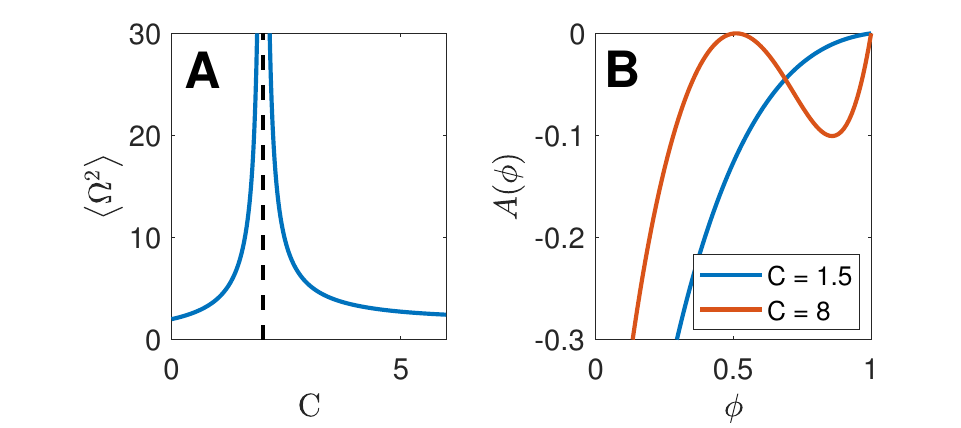}
\par\end{centering}
\caption{\label{fig:Omega2_Kauffman}(A) $\left\langle \Omega^{2}\right\rangle $
for the Kauffman model. The dashed vertical line marks the transition
at $C=2$. (B) The function $A(\phi)$ for two values of $C$, below
and above the transition. It shows the saddle point contributions
to $\left\langle \Omega^{2}\right\rangle $ at $\phi=1$ for both
values of $C$, and for $C>2$ the contribution at $0<\phi^{*}<1$.}
\end{figure}

Since $q(N_{A}/N)$ depends on $N_{A}$, the sum in Eq. \ref{eq: second-moment short sum}
is not trivial, so we will approximate it using the saddle point method.
At large $N$, the terms in the sum are equal to $\exp[NA(\phi)]$
within exponential accuracy, with
\[
A(\phi)=\phi\ln q+\left(1-\phi\right)\ln\left(1-q\right)-\phi\ln\phi-\left(1-\phi\right)\ln\left(1-\phi\right).
\]
This function has either one or two maxima, depending on whether $C<2$
or $C>2$, as Fig. \ref{fig:Omega2_Kauffman} illustrates. The sum
giving $\left\langle \Omega^{2}\right\rangle $ is dominated by the
contributions near these two maxima,
\[
\left\langle \Omega^{2}\right\rangle =\left\langle \Omega^{2}\right\rangle _{\text{near}}+\left\langle \Omega^{2}\right\rangle _{\text{far}},
\]
omitting the second term if there is only one maximum. $A(\phi)$
always has one maximum at $\phi=1$ with $A(1)=0$. i.e. the contribution
is of order 1. As $\phi$ represents the overlap between the two fixed
points $x$ and $y$, around this maximum they are similar to each
other; this is the reason to call this contribution $\langle\Omega^{2}\rangle_{\mathrm{near}}$.
Let them differ on $n\equiv N-N_{A}\ll N$ sites. Plugging into Eq.
\ref{eq:secondmoment-sum} gives:

\begin{equation}
\left\langle \Omega^{2}\right\rangle _{\text{near}}\approx1+\sum_{n=1}^{\infty}\frac{\left(nC/2\right)^{n}}{n!}e^{-nC/2}=\frac{1}{1+W\left(-\frac{C}{2}e^{-\frac{C}{2}}\right)}\ .\label{eq:Kauffman_inf_sum}
\end{equation}
In this expression, $W$ is again the Lambert function. This analytical
expression can be derived using a series given in Appendix \ref{sec:Lambert_W}.

The second maximum, that exists in the fluctuating phase $C>2$, turns
out to be at $\phi^{*}$ that satisfies
\[
\phi^{*}=q(\phi^{*})\,,
\]
so $\phi^{*}$ is precisely the fixed point of the mean-field evolution
of the overlap fraction, Eq. (\ref{eq:q_definition}). This can be
deduced by solving $A'(\phi)=0$ but there is a simpler derivation
that applies beyond the Kauffman model that is given in Sec. \ref{subsec:Dynamics_Fixed_Points}.

At this extremum, as in the maximum at $\phi=1$, $A(\phi^{*})=0$
and the contribution to $\left\langle \Omega^{2}\right\rangle $ is
$O(1)$. The sum near this extremum can be approximated by an integral\textendash the
usual Gaussian saddle point approximation is sufficient. This, with
the help of $\phi^{*}=q(\phi^{*})$, whose solution is Eq.(\ref{eq:Kaufmann_phi_star_expression}),
gives
\begin{equation}
\left\langle \Omega^{2}\right\rangle _{\text{far}}=\frac{1}{\left|1-q'(\phi^{*})\right|}=\frac{1}{1-C(\phi^{*}-1/2)}=\frac{1}{1+W\left(-\frac{C}{2}e^{-\frac{C}{2}}\right)}\ .\label{eq:dift_clusters_Kauffman}
\end{equation}
(See \ref{subsec:Dynamics_Fixed_Points} for the derivation.) Fig.
\ref{fig:Omega2_Kauffman} shows the final result for $\left\langle \Omega^{2}\right\rangle $.
In both phases, $\left\langle \Omega^{2}\right\rangle $ is finite,
namely does not scale with $N$, except at $C=2$ where this number
diverges.

\subsubsection{The second moment and the organization of FPs in configuration space}

Consider now whether some of the fixed points of a given random cellular
automaton of the Kauffman model are similar to one another. This can
be investigated by considering the expression for $\langle\Omega^{2}\rangle$.
In the frozen phase, $C<2$, the only maximum of $A(\phi)$ is at
$\phi=1$, and the only contribution to $\left\langle \Omega^{2}\right\rangle $
is $\left\langle \Omega^{2}\right\rangle _{\text{near}}.$ Thus, all
pairs of FPs are at a finite distance from each other (finite number
of sites where $x_{i}\ne y_{i}$). The expression in Eq. (\ref{eq:Kauffman_inf_sum})
simplifies (see Appendix \ref{sec:Lambert_W}) to
\begin{equation}
\left\langle \Omega^{2}\right\rangle =\frac{1}{1-C/2}\ \ \ \ \ \ \ \ (C<2)\label{eq:Omega2_Kauffman_frozen}
\end{equation}
This simple result is derived in a different way in Sec. \ref{subsec:short-cycleFPS}.

Throughout the fluctuating phase, $C>2$, the contributions to $\left\langle \Omega^{2}\right\rangle $
are non-zero from both $\left\langle \Omega^{2}\right\rangle _{\text{near}}$
and $\left\langle \Omega^{2}\right\rangle _{\mathrm{far}}$, so there
are pairs of fixed points at a finite distance or at an extensive
distance from each other. In other words, FPs are arranged in clusters
of fixed points at a finite distance from each other, while fixed
points in different clusters are at an extensive distance from each
other. Both contributions are the same by Eqs. (\ref{eq:Kauffman_inf_sum}),
(\ref{eq:dift_clusters_Kauffman}), giving

\[
\langle\Omega^{2}\rangle=\frac{2}{1+W\left(-\frac{C}{2}e^{-\frac{C}{2}}\right)}\ .
\]
Approaching the transition from above, $C\to2^{+}$, $\left\langle \Omega^{2}\right\rangle \approx\frac{2}{C/2-1}$.
So $\left\langle \Omega^{2}\right\rangle $ diverges with the same
exponent as for $C\to2^{-}$, but with a factor of 2 difference in
the prefactor.

When $C$ is close to 2, the sum in Eq. (\ref{eq: second-moment short sum})
is dominated by values of $N_{A}$ that are close to $N$, but $N-N_{A}$
is not of order 1 as it is when $C$ is below 2 by a finite amount.
Thus, we write $N_{A}=N(1-u)$ and find that Eq. (\ref{eq: second-moment short sum})
becomes, for $u$ small and $C$ close to 2:

\[
\langle\Omega^{2}\rangle\simeq\frac{\sqrt{N}}{2\pi}\int_{0}^{1}\frac{du}{\sqrt{u}}\exp\left[-\frac{Nu(C-2-2u)^{2}}{8}\right]\simeq N^{\frac{1}{3}}f(N^{\frac{1}{3}}(C-2))
\]
where $f(z)=\int_{0}^{\infty}\frac{dv}{\sqrt{2\pi v}}\exp\left[-\frac{v(z-2v)^{2}}{8}\right]$.
In particular, at the transition, i.e., for $z=0$, one gets $\langle\Omega^{2}\rangle=\frac{2^{\frac{2}{3}}}{\sqrt{\pi}}\Gamma(\frac{7}{6})N^{\frac{1}{3}}\simeq.83N^{\frac{1}{3}}$.
This result is correct whether $C$ is above or below 2, and one can
see that it matches the $C-2$ dependence found above outside the
critical region.

\subsection{Distribution of number of FPs in the frozen phase, based on short
cycles\label{subsec:short-cycleFPS}}

In Sec. \ref{subsec:Kauffman-model_moments} we saw that in the frozen
phase, $C<2$, all fixed points are at a finite distance from each
other. We also found an expression for $\left\langle \Omega^{2}\right\rangle $
as an infinite sum, which reduced to $1/\left(1-C/2\right)$, see
Eq. (\ref{eq:Omega2_Kauffman_frozen}). Here we explain the origin
of these similar fixed points, and obtain the full distribution $P(\Omega)$
in the frozen phase.

\subsubsection{The graph structure of sites differing between FPs}

\begin{figure}
\begin{centering}
\includegraphics[width=0.45\textwidth]{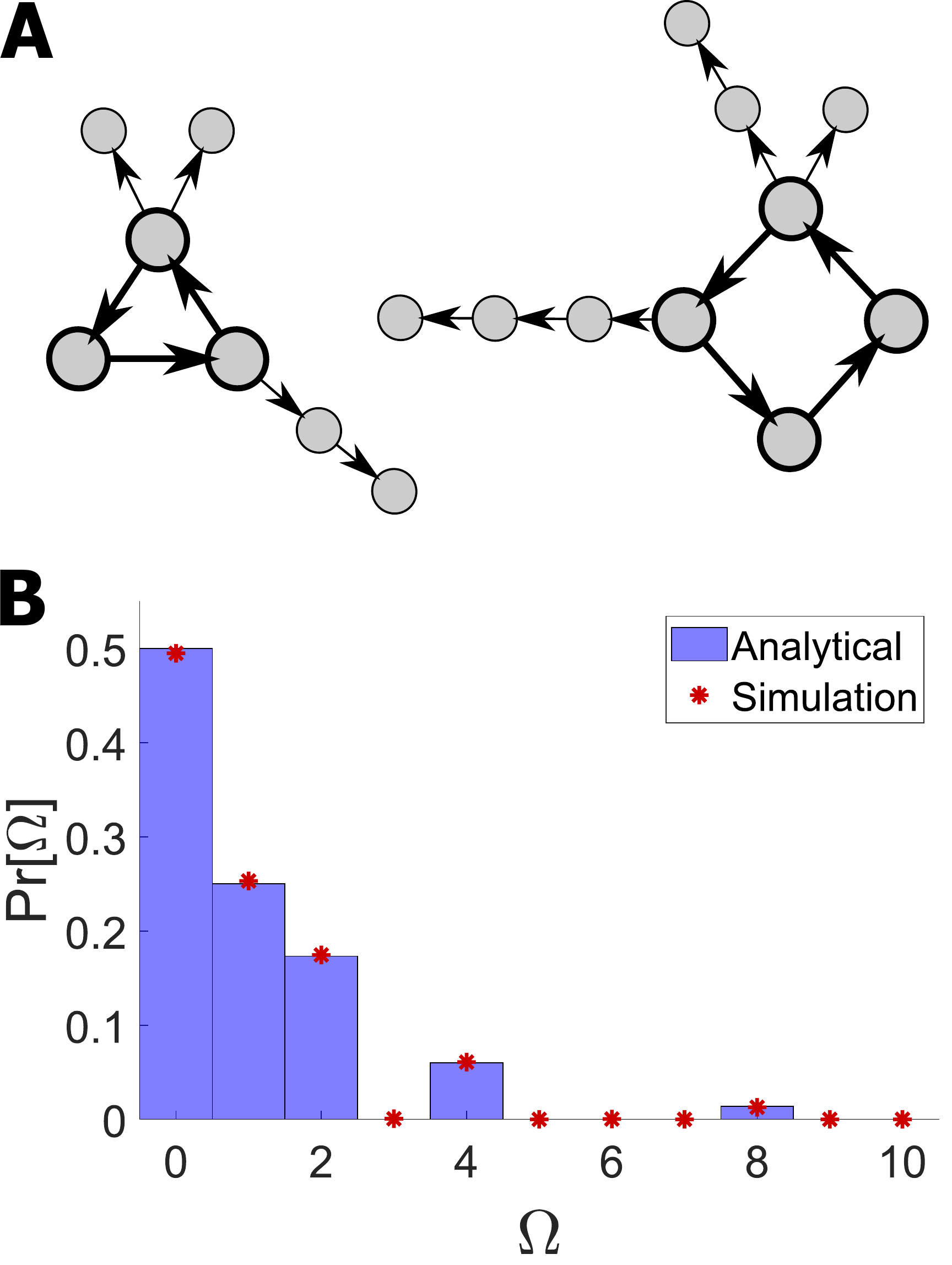}
\par\end{centering}
\caption{\label{fig:P_Omega_frozen}\textbf{Distribution of the number of FPs
in the frozen phase.} (A) Example of the subset $I$, of sites on
which assignments change between FPs in the frozen phase. (B) The
final $P(\Omega)$, theory (blue bars) against exhaustive searches
after graph simplification (red points). Both panels are obtained
by first applying the simplification algorithm (see main text) for
instances with $C=1.5$ and $N=5000$ sites.}
\end{figure}

Suppose we have two FPs, $x,y$, that differ from each other on a
finite subset $I$ of sites. The graph on $I$ has a special structure,
illustrated in Fig. \ref{fig:P_Omega_frozen}(A). To see the structure,
start from some site $i\in I$. Since $x_{i}\ne y_{i}$, there must
be a site $j\rightarrow i$ for which $x_{j}\ne y_{j}$, meaning that
$j\in I$. This way, one can continue to go upstream, following the
inputs to $j$ and so on, to create a string of sites in $I$. Since
$I$ is finite, this string must close on itself after a finite number
of steps, so there must be a finite directed cycle in $I$. In random
sparse graphs, cycles of finite length are distant from each other
(the distance between two finite cycles increases with $N$ \citep{bollobas_random_2001}),
and so the connected component of site $i$ can include only a single
cycle, possibly with trees emanating from the sites on the cycle.
There may be a finite number of such connected components of $I$.
Thus $I$ has a specific structure: it is made up of a finite number
of cycles with trees emanating from them.

In the frozen phase, any two FPs were found above to be at a finite
distance from one another, so the set where two such FPs differ always
has the structure of disconnected cycles with emanating trees. The
fact that in this phase any two distinct FPs are only different on
a finite number of sites, with this connectivity, suggests that for
any random Kauffman network there is a specific subset of sites which
can differ between different FPs, also made up of disjoint cycles
with trees emanating from them, while all other sites are the same
in all FPs.

Now we can justify this last assumption. First, we show that for typical
sites, that are located in a locally tree-like neighborhood, the value
that they may take in an FP is unique. We first provide numerical
evidence for this. We use an algorithm that simplifies an instance
of the Kauffman model, iteratively finding sites and edges that have
the same value in all FPs and removing them. We then analyze this
algorithm's dynamics to show that indeed most sites have the same
values in all FPs. (They form the ``stable core'', \citep{flyvbjerg1988,flyvbjerg1989,Bastolla1997,Kaufman2005,mihaljev2006}.)
The algorithm makes the following modifications to the system:

(1) The value $x_{i}$ of a site $i$ that has no inputs has a unique
value, in all FPs. The site can then be removed, along with the edges
outgoing from it. The functions downstream are reduced to functions
of fewer variables, by fixing the value of $x_{i}$.

(2) When there is an edge $j\to i$ such that the function $f_{i}$
giving $x_{i}$ does not depend on the value of $x_{j}$, the edge
can be removed.

These modifications do not change the number of fixed points of the
system. They are applied alternately until there are no vertices or
edges to which they can be applied. Running this algorithm shows that
instances of large systems can be reduced significantly, and for large
$N$ the outcome is of the form described above for $I$: a few cycles,
with downstream trees emanating from the cycles, see Fig. \ref{fig:P_Omega_frozen}(A).
The difference $I$ between two FPs will include some or all of the
connected components found this way. The simplified graphs can be
exhaustively searched to count FPs, which is used to find the probability
distribution for $\Omega,$ and thus to test the theory below. The
structure in Fig. \ref{fig:P_Omega_frozen}(A) is reminiscent of the
fluctuating components in the dynamics, Fig. \ref{fig:Dynamics}(C);
however, here we consider a different object, namely the remaining
nodes after the pruning algorithm.

The analysis of the algorithm proceeds by defining auxiliary dynamics
describing how the number of sites that are removed changes at each
step. In this dynamics, each site takes the values $0$ or $1$ once
its value has been determined, and $u$ when its value is yet unknown,
or varies between FPs. It is shown that in the frozen phase the fraction
of sites with value $u$ goes to zero when $t\rightarrow\infty$.
The detailed analysis is given in Appendix \ref{sec:frozen-sites}.
This suggests that all but $O(1)$ sites have uniquely determined
values (which is seen in the simulations). If a site remains in the
simplified graph, at least one of its inputs must also be in the simplified
graph. As for $I$ it follows that the simplified graph is made up
of finite cycles plus sites downstream from them.

\subsubsection{The number of FPs on a finite cycle}

We can now count the number of fixed points, by considering only the
number of FPs in the simplified graph. We will find that it is related
to the number of short cycles of certain types. (Ref. \citep{aracena2008}
found bounds on the number of fixed points in the fluctuating phase
using cycles in a similar way.) Although the simplified graph is made
up of cycles with trees emanating from them, we only need to evaluate
the number of FPs that can be assigned to the cycles, as once the
values on the cycle are given, the assignments of the downstream trees
is uniquely determined.

For a cycle of length $n$, denote by $i_{1}\to i_{2}\to...\to i_{n}\to i_{1}$
the indices of the sites along the cycle. After the reduction of the
system described in the previous section, there are reduced functions
$x_{i_{k}}=f_{i_{k}}^{\text{red}}\left(x_{i_{k-1}}\right)$ that must
be satisfied at an FP. They are random functions of one binary variable,
since they are obtained by reducing the original many-variable functions
by fixing all but one input of the function. The inputs that are fixed
come from random sites in the system, that are therefore almost surely
not in the simplified graph, and therefore have fixed values 0 or
1 with equal probability and independently between them.

\begin{figure}
\centering{}\includegraphics[width=0.5\textwidth]{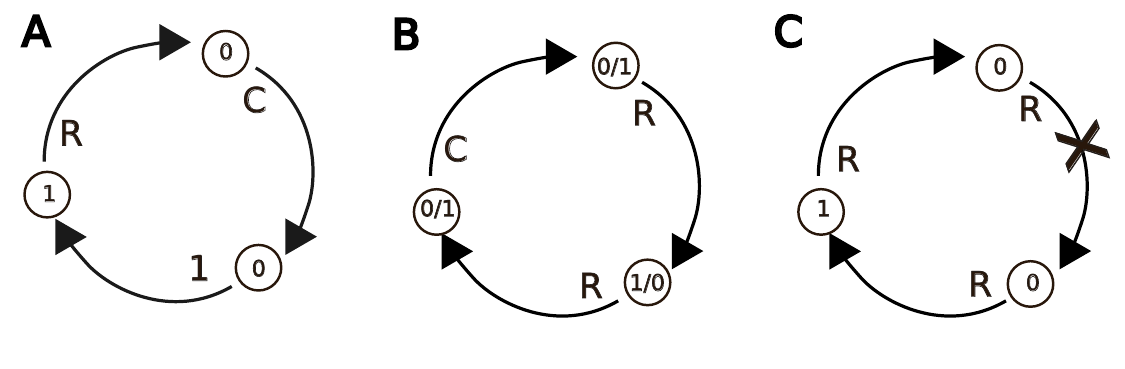}\caption{\label{fig:Examples-of-possible}Examples of cycles that may affect
the FP number in the Kauffman model. The functions are labeled on
the arrows: The constant function $f(x)=1$ is labeled as 1; the ``copy''
function $f(x)=x$ as ``C'', and the ``reverse'' function $f(x)=1-x$
as ``R.'' (A) A cycle with any constant function has a unique fixed
point and is removed from the reduced graph: in this example the function
1 is contant so, starting from it, all the sites along the cycle are
determined. (B) A cycle with two fixed points. (C) A cycle with no
consistent fixed points.}
\end{figure}

There are several cases depending on what the functions along the
cycle are, see Fig. \ref{fig:Examples-of-possible}. First, notice
that the cycles cannot have any reduced function that is a constant,
as in Fig. \ref{fig:Examples-of-possible}A. If one of the $\left\{ f_{i_{k}}^{\text{red}}\right\} $
is a constant function, then there is a unique FP assignment to the
cycle: because $f_{i_{k}}^{\mathrm{red}}$ is a constant function,
the value at site $i_{k}$ is fixed, and from there the rest of the
sites are fixed as well by following the cycle. In this case, the
cycle has uniquely determined values and it is not in the simplified
graph.

For the cycles that do belong to the reduced graph, there are either
0 or 2 fixed-point assignments. To find them, one can start by assigning
$x_{i_{1}}^{*}=1$, and from there obtaining the values $x_{i_{2}}^{*},..x_{i_{n}}^{*}$
along the cycle. There are no assignments if $x_{i_{1}}^{*}\ne f_{i_{1}}^{\text{red}}\left(x_{i_{n}}^{*}\right)$
(that is, if the cycle does not ``close on itself''); the probability
of this is $1/2$. Otherwise there are two FP assignments because
the values on the entire cycle can be flipped, as all functions are
either ``copy'' or ``reverse''. So the number of FPs for the network
is zero if the reduced graph has any cycle without FPs; and otherwise,
the number of FPs is $2^{k}$, where $k$ is the number of cycles
in the reduced graph. Thus determining the full distribution for $\Omega$
reduces to finding the distribution of $k$ and the probability to
have a cycle with 0 fixed points.

\subsubsection{The full FP number distribution in the frozen phase}

We consider a cycle in the graph describing the system, of length
$n$. What is the probability that it is in the reduced graph, and
that it has 0 or 2 fixed points? Every site that gives an input to
the cycle will be eliminated during the reduction process\footnote{A site that is not eliminated must have a cycle upstream and close
to it, so such a site would have to be close to two cycles. This does
not occur in a sparse random graph.}, so the values of these inputs are fixed. So we can consider the
cycle with all the reduced one-variable functions along it. If any
function is constant, the cycle will also be eliminated. Each function
along a cycle is sampled with equal probability from the four functions
$\left\{ 0,1\right\} \rightarrow\left\{ 0,1\right\} $, two of which
are constant, so the probability that there are no constant functions
is $2^{-n}$. Supposing that this is the case, the conditional probabilities
for 0 or 2 fixed points are both 1/2. So in total the probabilities
are
\[
P\left(\Omega_{n}=0\right)=P\left(\Omega_{n}=2\right)=2^{-n-1}\ ,
\]
where $\Omega_{n}$ represents the number of FPs for this cycle. These
alternatives for different cycles are independent as the cycles are
far from each other.

Now in terms of $C,$ the mean number of directed cycles of length
$n$ is $\lambda_{n}=C^{n}/n$, and this number is Poisson distributed
\citep{bollobas2011random}. (This also applies to a graph where the
number of incoming edges is fixed to be exactly $C$ and the number
of outgoing edges may fluctuate, as the Kauffman model is often defined;
hence the results below also apply to this version of the Kauffman
model.) If there is at least one short cycle in the reduced graph
with no FP assignment, then there is no FP assignment for the entire
system. The probability that all cycles of length $n$ \emph{do }have
a FP assignment is
\begin{align*}
\sum_{j=0}^{\infty}\frac{e^{-\lambda_{n}}}{j!}\lambda_{n}^{j}\left(1-2^{-n-1}\right)^{j} & =e^{-\left(C/2\right)^{n}/\left(2n\right)}
\end{align*}
where $j$ represents the number of cycles of length $n$. So the
probability that all cycles of \emph{any} length $n$ have FP assignments
is
\begin{align*}
\text{Pr}\left[\Omega\ne0\right] & =\prod_{n=1}^{\infty}e^{-\left(C/2\right)^{n}/\left(2n\right)}=\left(1-C/2\right)^{1/2}\,.
\end{align*}
This is the probability that there is at least one FP of the system.

Now to determine the distribution of the number of fixed points, we
must consider how many cycles allow different choices. The number
of cycles of length $n$ with 2 FP assignments is Poisson distributed
with mean $\lambda_{n}P\left(\Omega_{n}=2\right)$, and so $k$, the
total number of cycles that have 2 FP assignments, is Poisson distributed
with mean
\[
m_{2}=\sum_{n=1}^{\infty}\lambda_{n}P\left(\Omega_{n}=2\right)=\sum_{n=1}^{\infty}\frac{C^{n}}{n}2^{-n-1}=-\frac{1}{2}\ln\left(1-C/2\right)
\]
When there are $k$ such cycles, and no cycles without an FP, there
are $2^{k}$ FPs of the entire system.

This gives the probability distribution: The probability for $2^{k}$
fixed points is
\begin{align*}
\Pr\left[\Omega=2^{k}\right] & =\text{Pr}\left[\Omega\ne0\right]\Pr\left[k\text{ cycles with 2 FPs}\right]\\
 & =\left(1-C/2\right)^{1/2}\frac{m_{2}^{k}}{k!}e^{-m_{2}}=\frac{1-C/2}{2^{k}k!}\left[\ln\left(\frac{1}{1-C/2}\right)\right]^{k}
\end{align*}
where in the first equality the probability factorizes due to properties
of Poisson distributions\footnote{For each cycle length $n$, the number of cycles with 0 or 2 FPs are
independent Poisson random variables. Therefore, the number of cycles
of any length with zero FPs, which sets $\text{Pr}\left[\Omega\ne0\right]$,
is independent of the number of cycles with 2 FPs, which sets $\Pr\left[k\text{ cycles with 2 FPs}\right]$.}. Together

\begin{equation}
P\left(\Omega\right)=\begin{cases}
1-\left(1-C/2\right)^{1/2} & \Omega=0\\
\left(1-C/2\right)\frac{1}{2^{k}k!}\left[\ln\frac{1}{1-C/2}\right]^{k} & \Omega=2^{k},k\ge0\\
0 & \mathrm{otherwise}
\end{cases}\label{eq:probability distribution}
\end{equation}
This distribution is in excellent agreement with results of an exhaustive
search for FPs, see Fig. \ref{fig:P_Omega_frozen}(B), in which the
simplifying algorithm described above is applied to the graph and
then all fixed points are found using an exhaustive search on the
simplified graph.

From $P(\Omega)$, the $r$-th moment can be calculated:
\begin{equation}
\left\langle \Omega^{r}\right\rangle =\left(1-C/2\right)^{1-2^{r-1}}\label{eq:frozen_phase_moments-1-1}
\end{equation}
which agrees with the calculation above for $\left\langle \Omega\right\rangle $
and $\left\langle \Omega^{2}\right\rangle $ in the frozen phase,
Eq. (\ref{eq:Omega2_Kauffman_frozen}).

Note that, approaching the transition at $C=2$ from below, one obtains
that $\left\langle \Omega^{r}\right\rangle $ for $r\ge2$ diverges
while the first moment remains constant at the transition, while the
probability of having an FP goes to zero as $P\left(\Omega>0\right)\sim(2-C)^{1/2}$.
This is possible because the number of fixed points grows exponentially
with the number of cycles. The value of $\langle\Omega^{r}\rangle$
is dominated by contributions where the number of cycles with two
possible fixed point assignments is $2^{r-1}\log\frac{1}{2-C}$. This
is very unlikely (typically there are about $\log\frac{1}{C-2}$ cycles),
but the number of fixed points is very big when it happens.

\subsubsection{The structure of FPs in the fluctuating phase}

Some of the ideas just considered are also helpful for understanding
fixed points in the fluctuating phase, although they do not lead to
the full distribution function. The two contributions to $\langle\Omega^{2}\rangle$,
namely $\langle\Omega^{2}\rangle_{\mathrm{near}}$ and $\langle\Omega^{2}\rangle_{\mathrm{far}},$
count pairs of fixed points with a full overlap or a partial overlap
of $\phi^{*}$ respectively. These contributions can be understood
by partitioning fixed points into clusters, where each cluster is
made of fixed points that differ from each other only at a finite
number of points, on and near cycles. $\Omega^{2}$ equals the number
of pairs of fixed points (counting a pair $\left(i,j\right)$ and
$\left(j,i\right)$ separately and also counting $\left(i,i\right)$
as a pair). If there are $p$ clusters, and $n_{i}$ fixed points
in the $i^{\mathrm{th}}$ cluster, then $\Omega^{2}=\sum_{i=1}^{p}n_{i}^{2}+2\sum_{i=1}^{p}\sum_{j=1}^{i-1}n_{i}n_{j}\equiv\left(\Omega^{2}\right){}_{\mathrm{near}}+\left(\Omega^{2}\right){}_{\mathrm{far}}$.
The means of the two terms give the contributions $\langle\Omega^{2}\rangle_{\mathrm{near}}$
and $\langle\Omega^{2}\rangle_{\mathrm{far}}$ found above, since
$\langle\Omega^{2}\rangle_{\mathrm{near}}$ represented the part of
the sum (\ref{eq: second-moment short sum}) near the saddle point
with $\phi=1$, corresponding to pairs of fixed points that overlap
at almost all the sites, while the latter includes the rest of the
pairs of fixed points.

In the frozen phase, there is only one cluster, and above we were
able to find the full distribution function for the number of fixed
points in this cluster. In the fluctuating phase, the fixed points
in each cluster still differ from one another by the values along
cycles and trees extending from the cycles (for the same reason as
in the frozen phase). However, it is not easy to find \emph{which
}cycles can be changed in this way. This is because the reduced version
(using the reduction algorithm from above) of the graph contains an
extensive number of sites and thus many cycles,which can be close
to one another or overlap. Nevertheless, since $\langle\Omega\rangle$
is finite, there must be only finitely many FPs in each cluster. The
reason is probably that there are only finitely many cycles that can
be changed without destabilizing the whole FP. We also do not know
the full probability distribution for the number of distinct clusters.
Because these phenomena are harder to understand, we do not know the
full probability distribution in the fluctuating phase.

The calculation of $\langle\Omega^{2}\rangle$ gives one interesting
result about the clusters: since $\langle\Omega^{2}\rangle_{\mathrm{far}}$
is dominated by one saddle point, where all pairs of fixed points
overlap at a fraction $\phi^{*}$ of the sites, any two fixed points
from distinct clusters have the same distance from one another, $1-\phi^{*}$.

\section{Different types of transitions, and how they affect the fixed points\label{sec:Mean-magnetization}}

We will now consider the other models shown in the box above, which
exhibit a range of phenomena. These models can be understood more
intuitively if we consider sites at 0 and 1 as being ``off'' and
``on'' respectively. To derive their properties, we will use both
of the methods from the previous section, the geometrical method based
on cycles (which gives a full distribution function in the frozen
phases) and the evaluation of moments, which gives less complete information,
but works in any phase and makes an interesting connection between
the dynamics and the number of fixed points. We will first discuss
this connection.

We consider mostly the first moment. In the Kauffman model, the first
moment is very simple (always equal to 1), so we analyze mainly the
second moment.

\subsection{Formula for Fixed Points based on Dynamics\label{subsec:Dynamics_Fixed_Points}}

Here we will give a formula for the moments of $\Omega$ in any general
cellular automaton. Consider a model with binary variables, with random
sparse connections as above. For an assignment $x$ of values for
the $N$ sites, let $\psi$ be the fraction of sites for which $x_{i}=1$,
which we will below call the magnetization. The time dependence of
$\psi_{t}$ is described by a dynamical equation when $N$ is large,
\begin{equation}
\psi_{t+1}=q\left(\psi_{t}\right),\label{eq:mag_MF_dynamics}
\end{equation}
aside from small fluctuations. Here $q$ is some function that can
be obtained for each model, given its ensemble of functions. Notice
that $q(\psi_{t})$ can also be interpreted as the probability that
a random function from the ensemble takes 1 as its value for inputs
chosen randomly from among all the sites.

The first moment $\langle\Omega\rangle$ can be related to $q$, (see
also Ref. \citep{samuelsson_superpolynomial_2003}) following a derivation
similar to that in Sec. \ref{subsec:Kauffman-model_moments}, see
Appendix \ref{sec:Saddle-points_from_MF_dynamics}. Begin with Eq.
(\ref{eq:mean_omega_as_sum}). The sum over $D$ is the probability
that $x$ is a fixed point. If the value of a site is equal to 1,
then the probability that it is fixed (for one step of evolution)
is $q(N_{1}/N)$$,$ because the condition that it is fixed means
that the function assigned to it comes out to equal 1. Similarly,
if its value is 0, the probability that this configuration is fixed
at the site is $1-q(N_{1}/N)$ . Thus the expected number of fixed
points is 
\begin{equation}
\langle\Omega\rangle=\sum_{N_{1}=0}^{N}\binom{N}{N_{1}}q(N_{1}/N)^{N_{1}}(1-q(N_{1}/N))^{N-N_{1}}.\label{eq:firstmoment-qsum}
\end{equation}
To evaluate this sum by the saddle point method, we write the terms
to exponential accuracy as $e^{A(\psi)N}$, where $\psi=N_{1}/N$
and $A(\psi)=\psi\ln\frac{q(\psi)}{\psi}+(1-\psi)\ln\frac{1-q(\psi)}{1-\psi}$.
Solving $A'(\psi)=0$ algebraically, in order to find the saddle points,
is not possible for a general function $q$; however following \citep{samuelsson_superpolynomial_2003},
we can argue that as $-\ln x$ is convex, Jensen's inequality implies
$A(\psi)\leq\ln[\psi\frac{q(\psi)}{\psi}+(1-\psi)\frac{1-q(\psi)}{1-\psi}]=0$.
This maximal possible value, $A=0$, is attained only when $\frac{q(\psi^{*})}{\psi^{*}}=\frac{1-q(\psi^{*})}{1-\psi^{*}}$,
or $\psi^{*}=q(\psi^{*})$, where $\psi^{*}$ is the position of the
maximum\footnote{There can be local maxima with smaller values of $A$, which would
not be fixed points of $q$. Their contributions to $\langle\Omega\rangle$
can be neglected compared to contributions near the points where $A=0$,
if there are any such points. In fact such points always exist because
$q$ is continuous and maps the interval $[0,1]$ into itself, which
implies that $q(\phi)=\phi$ has a solution.}. Such a value is a fixed point of the dynamics of the \emph{magnetization}
(a different concept from the fixed points that we have been studying,
which are fixed point \emph{configurations} of the automaton). It
makes sense that the main contributions to $\langle\Omega\rangle$
come from fixed points of $q$, since in order for a configuration
to be an FP, i.e., for \emph{each} site's value not to change in time,
the magnetization $\psi$ must a fortiori be independent of time,
which means that it is a fixed point of $\psi\rightarrow q(\psi)$.
As $q(\psi)$ is a function $\left[0,1\right]\rightarrow\left[0,1\right]$,
this must have at least one solution.

Since $A(\psi^{*})=0$ at the extrema and the contribution to the
sum is $e^{A(\psi)N}$ (apart from possibly a power of $N$), the
number of fixed points neither grows nor decays exponentially with
$N$. The sum of Eq. (\ref{eq:firstmoment-qsum}) in the vicinity
of an extremum can be approximated by a Gaussian integral when $\psi^{*}\notin\left\{ 0,1\right\} $,
giving
\begin{equation}
\frac{1}{|1-q'(\psi^{*})|}\ ;\label{eq:saddlepoint-1}
\end{equation}
see Appendix \ref{sec:Saddle-points_from_MF_dynamics}. In particular
the expected number of fixed points approaches a finite limit when
$N\to\infty$, unless $q'(\psi^{*})=1$. (The formula for the \textit{second
moment }in the Kauffman model has the same form (Eq. (\ref{eq: second-moment short sum}))
but this is special for the Kauffman model because in it the sum for
the second moments, Eq. (\ref{eq:secondmoment-sum}), can be replaced
by a sum over just the overlap fraction, Eq. (\ref{eq: second-moment short sum}),
that has the same form as the general expression for the first moment.
In the appendices the more general sums are evaluated.)

The second moment $\langle\Omega^{2}\rangle$ can be calculated in
a similar way from Eq. (\ref{eq:secondmoment-sum}), with the functions
$q_{00},\dots,q_{11}$ describing the dynamics of two configurations.
That is, let $\psi_{00}=N_{00}/N$ etc. The functions $q_{00},q_{10},q_{01},q_{11}$
give the time dependence of these variables, e.g. $\psi_{00}(t+1)=q_{00}(\psi_{11}(t),\psi_{10}(t),\psi_{01}(t),\psi_{00}(t)).$
The extrema are solutions to simultaneous equations $\psi_{ij}=q_{ij}(\psi_{ij})$
where $i,j\in$\{0,1\}. The contribution to $\langle\Omega^{2}\rangle$
from an extremum away from the boundary of the domain of physical
values of $\psi_{ij}$ can be found by a Gaussian approximation, generalizing
Eq. (\ref{eq:saddlepoint-1}). There is always at least one solution:
this is when the two copies are the same, and $\psi_{01}=\psi_{10}=0$,
$\psi_{11}=\psi^{*}$, $\psi_{00}=1-\psi^{*}$.

The Gaussian approximation will break for $\langle\Omega\rangle$
for extrema at $\psi^{*}=0$ or $\psi^{*}=1$, and it will break for
$\langle\Omega^{2}\rangle$ for extrema on the boundary of the domain
of $\psi_{ij}$'s. In this case one has to evaluate a sum (much like
the contribution to $\langle\Omega^{2}\rangle_{\mathrm{near}}$ in
the Kauffman model), since then the terms of Eqs. (\ref{eq:secondmoment-sum})
or (\ref{eq:firstmoment-qsum}) are not smooth enough to be approximated
by integrals. The sums can still be expressed in terms of the derivatives
of $q$ at the saddle point (see Appendix \ref{sec:generalsum}),
and for some of the models studied here they can be expressed in terms
of the Lambert function like for the Kauffman model.

From this dynamical point of view, where the moments are written entirely
in terms of the function $q\left(\psi\right)$ and its analogues for
several copies of a system, there are different universality classes
for the behavior of $\Omega$ that are determined by the type of bifurcation
transition that occurs in the graphs of these functions, as described
in several examples in the remainder of this section.

We have seen that all fixed point configurations have a magnetization
that is very close to a fixed point of the magnetization dynamics,
$\psi^{*}=q(\psi^{*})$ (almost surely).This is intuitively reasonable,
but it allows something surprising to happen: If $\psi^{*}$ is an
\emph{unstable} fixed point of $q$ there are still FPs nearby. Since
unstable fixed points still have a finite value of $q'(\psi^{*})$
(one that is $>1$), Eq. (\ref{eq:saddlepoint-1}) shows that there
are $O(1)$ FPs associated with them. There are examples of this in
all the models considered below. For example, in the Inhibitory model,
Sec. \ref{subsec:The-Inhibitory-model}, simulations with parallel
dynamics show that in the non-frozen phase the solution to $\psi=q(\psi)$
is dynamically unstable towards a 2-cycle, yet the mean number of
FPs is not 0 because a contribution to $\langle\Omega\rangle$ is
produced by the unstable solution.

When the $\vec{\psi}^{*}$ at the saddle point does not have any components
that are equal to zero, the sum becomes smooth so it can be approximated
by an integral, and the integral is a Gaussian integral: The expression
has the form: $e^{NA(\vec{\psi})}$ times a coefficient that is a
power of $N$. Expanding gives $A=A(\vec{\psi}^{*})+\frac{1}{2}\frac{\partial^{2}A}{\partial\psi_{ij}\partial\psi_{rs}}(\psi_{ij}-\psi_{ij}^{*})(\psi_{rs}-\psi_{rs}^{*})$
because there is no first order term. The exponential decays when
the $n_{ij}$ change by an amount $\sim\sqrt{N}$, so there are many
terms in the sum over $n_{ij}$ within this range, and it can be approximated
by an integral. But when the maximum is at the edge of the range of
integration, it could be possible that the first derivative is nonzero,
and then the sum would decay on a scale of order 1, and it cannot
be approximated by an integral. (It turns out that $A$ does not even
have a Taylor series expansion around 0 in this case. See appendix
\ref{sec:Saddle-points_from_MF_dynamics}.)

\subsection{The Excitatory model\label{subsec:The-Excitatory-model}}

\begin{figure}
\begin{centering}
\includegraphics[viewport=10bp 30bp 395bp 450bp,clip,width=0.6\textwidth]{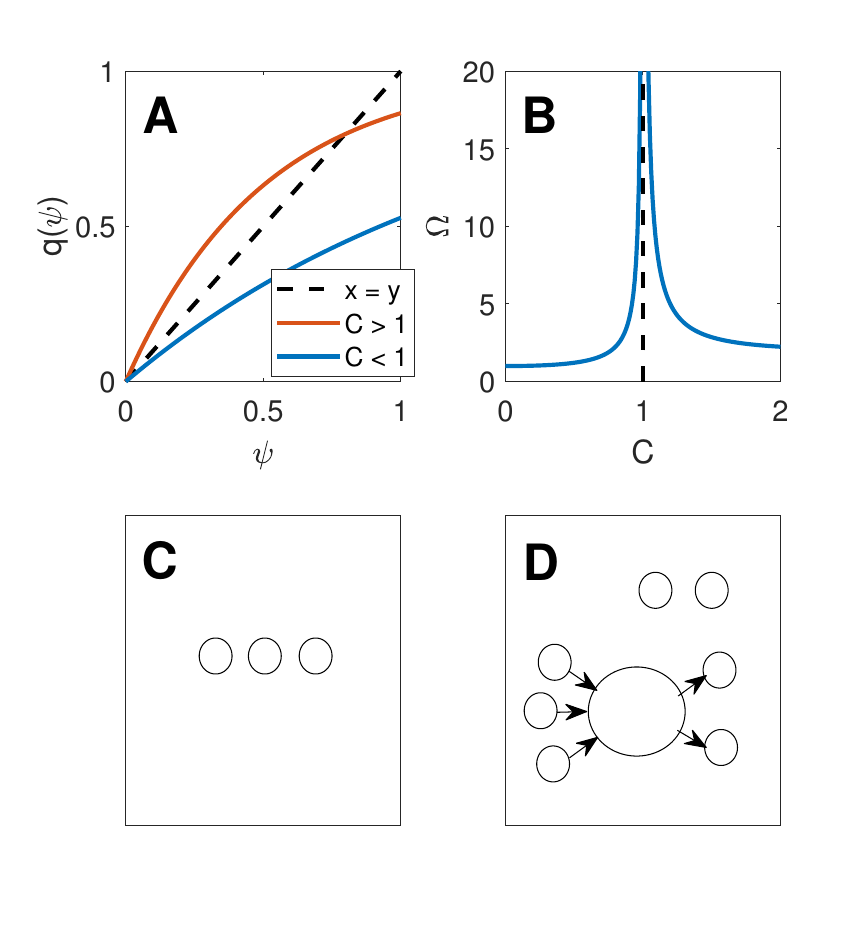}
\par\end{centering}
\caption{\label{fig:excitatory_model}\textbf{The excitatory model.} (A) The
function $q(\psi)$ for different values of $C$, above and below
$C_{\text{crit}}=1$. (B) The first moment $\left\langle \Omega\right\rangle $.
(C) SCCs in the phase $C<C_{\text{crit}}$. Each circle denotes an
SCC, each containing a finite number of sites. (D) SCCs in the phase
$C>C_{\text{crit}}$ include a giant SCC comprised of a finite fraction
of the sites (large circle), and possibly a finite number of SCCs
with a finite number of sites (small circles). They may be disconnected
from the giant component, or with directed paths to or from it. Arrows
denote paths of links between the SCCs.}
\end{figure}

The model of this section and the next have simpler dynamical transitions
than the Kauffman model: the mean-field dynamics changes rather than
the system going from a frozen to a fluctuating state with the same
mean-field values. This is reflected even in the first moment of the
number of fixed points. The excitatory model is defined by the dynamical
rule: $x_{i}(t+1)=1$ iff at least one incoming edge $j\to i$ has
$x_{j}(t)=1$. The magnetization $\psi$ satisfies $\psi_{t+1}=q(\psi_{t})$
with $q(\psi_{t})=1-e^{-C\psi_{t}}$. The value $\psi^{(1)}=0$ is
a fixed point of this equation, for all $C$. For $C>C_{\text{crit}}\equiv1$
there is another fixed point at some $\psi^{(2)}>0$.

The mean number of FPs $\left\langle \Omega\right\rangle $ diverges
on both sides of the transition $C_{\text{crit}}$, see Fig. \ref{fig:excitatory_model}(B).
The calculation is given in Appendix \ref{sec:excitatory_derivations}.
This can be seen from the graph of $q(\psi)$ in Fig. \ref{fig:excitatory_model}
(A). At $C_{\mathrm{crit}}$, where one fixed point bifurcates into
2, the line $y=x$ is tangent to the graph of $q$, so the slope is
1 and Eq. \ref{eq:saddlepoint-1} diverges. (This argument applies
actually only above $C_{\mathrm{crit}}$ to the nonzero saddle point,
$\psi^{(2)}$, since Eq. (\ref{eq:saddlepoint-1}) assumes $\psi^{*}\neq0,1$,
but the other contributions also diverge.)

This model is interesting because the FPs have a geometrical interpretation,
allowing one to understand and calculate $P(\Omega)$ on both sides
of the transition, and to understand why there are two clusters of
fixed points. At an FP, if there is an edge $j\to i$ then $x_{j}\le x_{i}$,
namely $x_{i}=1$ if $x_{j}=1$. This then applies to any path between
sites: if there is a path from $j$ to $i$, $j\to..\to i$, then
$x_{j}\le x_{i}$ at an FP; i.e., if a site must be on if any site
upstream of it is. Now consider the strongly connected components
(SCCs) of a directed graph. An SCC is a subset of vertices for which
there are directed paths in both directions between all vertices in
the subset. From the above, at an FP the value $x_{i}$ must be equal
for all sites $i$ in an SCC: the entire SCC is either off or on.
Next, SCCs admit a partial ordering: if there is a path from one SCC
to another (but not in both directions, otherwise they would be the
same SCC), for a site $j$ in the upstream SCC and a site $i$ in
the downstream one, there is a path $j\to..\to i$ and so at an FP
$x_{j}\le x_{i}$. In other words, if an SCC is on, then all of the
downstream SCCs will also be on.

The transition at $C_{\text{crit}}$ is due to the percolation transition
which changes the geometry of the SCCs, with the appearance of an
SCC giant component, see Fig. \ref{fig:excitatory_model}(C,D). Below
the transition there are only a finite number of SCCs, all of finite
size, which are precisely the short directed cycles in the graph,
and which are disconnected from one another (with probability 1).
Each short cycle can be assigned either 0 or 1, giving several fixed
points. All sites that are not downstream from a cycle must be off
at a fixed point\footnote{If a site is on at an FP, then at least one site upstream from it
must also be on. Iterating this gives a sequence of sites that are
on, and the sequence must close into a cycle eventually.}, and so these FPs will be associated with the magnetization $\psi^{(1)}=0$.
Above the transition there is also a giant SCC, comprised of a finite
\emph{fraction} of the sites. This means that there are two clusters
of FPs, corresponding to whether the giant component is off or on;
the latter type of fixed point is associated with the saddle point
$\psi^{(2)}>0$, since the number of sites with value 1 is extensive
for such a fixed point. This picture allows to calculate $P(\Omega)$,
see details in Appendix \ref{sec:excitatory_derivations}. The dynamics
of this model is simpler than in the Kauffman model because even when
$C>C_{\mathrm{crit}}$ it is not fluctuating; the dynamics just changes:
below $C_{\mathrm{crit}}$ every state is attracted to frozen states
where $\psi\approx\psi^{(1)}=0$ (the small SCCs may be on) while
above $C_{\mathrm{crit}}$ it is attracted to the states near $\psi^{(2)}$
with a giant SCC, which are also frozen states. In addition, there
are still fixed points near $\psi^{(1)}=0$, which have small basins
of attraction (as they require all SCCs upstream of the giant SCC
to be at 0).

\subsection{The Double excitatory model\label{subsec:The-Double-excitatory}}

The double-excitatory model is defined by the dynamical rule: $x_{i}(t+1)=1$
iff at least \emph{two} of the incoming edges have $x_{j}(t)=1$.
This captures situations where there is a barrier for a transition,
such as a voltage threshold for a neuron to fire.

\begin{figure*}
\begin{centering}
\includegraphics[width=0.5\textwidth]{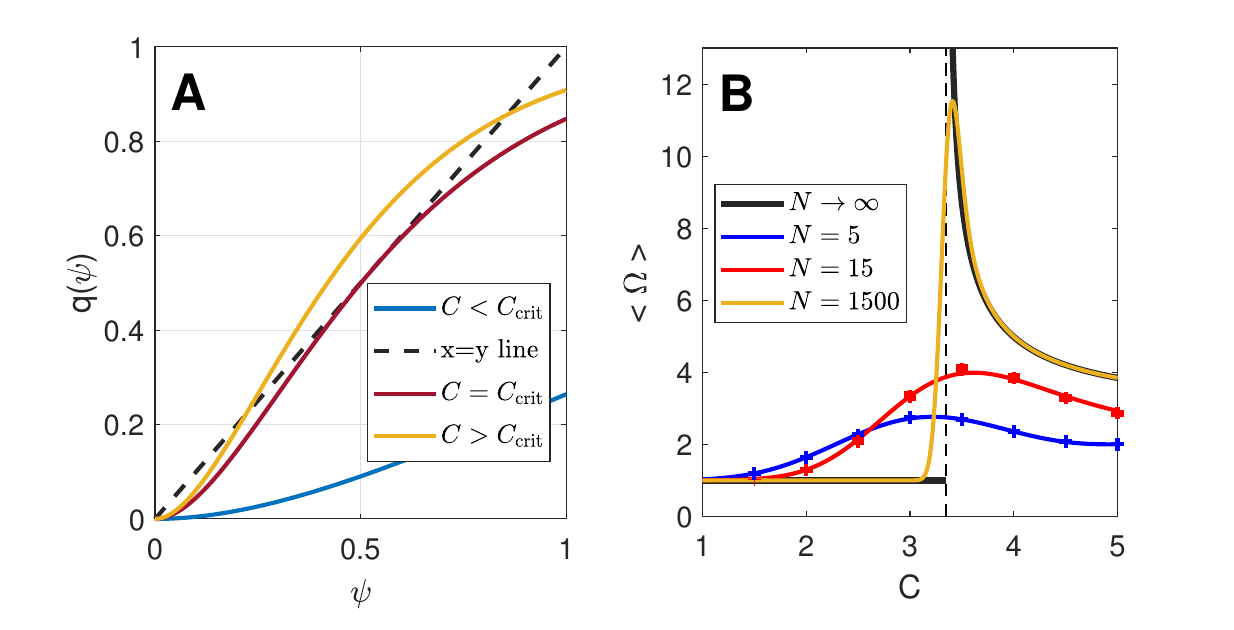}
\par\end{centering}
\caption{\label{fig:double-excitatory}\textbf{The double-excitatory model.}
(A) The saddle-point function for $\left\langle \Omega\right\rangle $
as a function of $\psi$, at $C=2,3.35,4$. (B) $\left\langle \Omega\right\rangle $
for the Double Excitatory model (black curve), and finite-$N$ behavior.
Solid curves are calculated from an exact evaluation of $\left\langle \Omega\right\rangle $
when expressed as a sum, and crosses (with 1SE error bars) are the
results of exhaustive searches at $N=5,15$, each averaged over 1000
disorder realizations. Dashed line: transition at $C_{\text{crit}}$.}
\end{figure*}

The magnetization $\psi$ satisfies $\psi_{t+1}=q(\psi_{t})$ with
$q(\psi_{t})=1-e^{-C\psi_{t}}\left(C\psi_{t}+1\right)$, see Fig.
\ref{fig:double-excitatory}(A). The value $\psi^{(1)}=0$ is a fixed
point for all $C$. For $C<C_{\mathrm{crit}}\approx3.35$, the solution
$\psi^{(1)}=0$ is unique. At $C_{\mathrm{crit}}$ there is a transition
where two additional fixed points appear, $\psi^{(2)}<\psi^{(3)}$.
This is a first order transition, in the sense that these fixed points
are not close to $\psi^{(1)}$ at $C_{\mathrm{crit}}$; the graph
of $y=q(x)$ has a convex part that first meets $y=x$ in a new pair
of points when $C$ increases above $C_{\mathrm{crit}}$ (see Fig.
\ref{fig:double-excitatory} A). $\psi^{(2)}$ is unstable, so the
dynamics reaches either of the stable FPs, $\psi^{(1)}$ or $\psi^{(3)}$,
depending on initial conditions. This transition is related to a transition
for an undirected graph, where a ``3-core'' appears when the average
degree of the vertices is increased (a subgraph where each vertex
has at least 3 neighbors)\citep{Pittel_1996}.

The first moment $\left\langle \Omega\right\rangle $ is calculated
with the same technique as above, see Appendix \ref{sec:Double-excitatory-derivations}.
The result is plotted in Fig. \ref{fig:double-excitatory}(B). At
$C<C_{\mathrm{crit}}$ just one term, exactly at $\psi^{(1)}=0$,
gives the total number of fixed points $\langle\Omega\rangle=1$.
(If $N_{1}>0$, the terms are smaller by at least $\frac{1}{N}$.)
The reason why there are no other FPs near the FP where all sites
are off, is that the double excitatory condition implies that each
site that is on belongs to at least two cycles of sites that are also
on. This cannot happen for a finite subset of the sites in an Erd\H{o}s\textendash Rényi
graph when $N\rightarrow\infty$. (This is also true for the random
graphs with multiple edges as defined in Sec. \ref{subsec:Kauffman-model_moments}.)
Above $C_{\mathrm{crit}}$ there are three contributions to $\left\langle \Omega\right\rangle $,
from the three solutions. In total, there is a divergence at $C\to C_{\mathrm{crit}}^{+}$.
This results in the discontinuous function shown in Fig. \ref{fig:double-excitatory}(B).
The finite-$N$ curves that approach this asymptotic behavior are
shown in Fig. \ref{fig:double-excitatory}(C), both from a numerical
summation of Eq. (\ref{eq:mean_omega_as_sum}) for finite $N$, and
from exhaustive searches for FPs at small $N=5,15$.

Close to and above the transition, $\left\langle \Omega\right\rangle \sim\frac{1}{\sqrt{C-C_{c}}}$,
see Appendix \ref{sec:Double-excitatory-derivations}. This can be
derived from the general behavior of $q(\psi)$ near the transition,
see Appendix \ref{sec:Double-excitatory-derivations}. The average
number of FPs also depends in a different way on the number of sites:
at $C=C_{c}$, $\langle\Omega\rangle\propto N^{\frac{1}{4}}$, whereas
in the excitatory model, where the transition is second order, $\langle\Omega\rangle\propto N^{\frac{1}{3}}$.

In the excitatory model, we found that the transition is connected
to the appearance of a subgraph that can be defined geometrically,
the large strongly connected component. The vertices that belong to
this can also be identified in a local way: by testing whether the
outward and inward trees leaving the vertices grow exponentially for
several generations or die off. In the double excitatory model, the
sites that are on at a FP with $\psi=\psi^{(3)}$ also seem to have
a geometrical interpretation, while those with $\psi=\psi^{(2)}$
seem not to, as suggested by the results of Ref. \citep{Johnson_2020}.

We have not calculated $\left\langle \Omega^{2}\right\rangle $, but
it will have the same qualitative features: For $C<C_{\mathrm{crit}}$,
there is exactly one FP; therefore $\left\langle \Omega^{2}\right\rangle =1$;
above the transition $\left\langle \Omega^{2}\right\rangle \ge\left\langle \Omega\right\rangle ^{2}$
and so there will be a divergence.

\subsection{The Inhibitory model\label{subsec:The-Inhibitory-model}}

The inhibitory model is defined by: $x_{i}(t+1)=1$ iff for all incoming
edges $x_{j}(t)=0$. It is of interest as a model of inhibitory interactions
in neural networks, and ecological interactions between species. There
is a phase transition in the magnetization dynamics at some value
$C_{\mathrm{crit}}$: $q(\psi)$ in the notation of Eq. (\ref{eq:mag_MF_dynamics})
always has one fixed point $\psi^{*}$, but at $C_{\mathrm{crit}}$
it becomes unstable to a 2-cycle. The change in the dynamics does
not lead to any singularity of $\langle\Omega\rangle$ since the fixed
point becomes unstable via $q'(\psi^{*})$ passing through $-1$,
so that Eq. (\ref{eq:saddlepoint-1}) is finite. On the other hand,
the second moment $\langle\Omega^{2}\rangle$ diverges at $C_{\mathrm{crit}}$,
and at $C>C_{\mathrm{crit}}$ there can be multiple clusters of FPs.
The derivation is detailed in Appendix \ref{sec:InhibitoryAppendix}.

The behavior of the moments is thus similar to the Kauffman model,
although unlike in the Kauffman model the transition has an effect
on the dynamics of the magnetization; this just does not lead to a
singularity of $\langle\Omega\rangle$. The calculations are more
complicated than in the Kauffman model, since the dynamics of two
copies cannot be reduced to just the dynamics of the overlap fraction
$\phi$.

Simulations show that in this case, the fluctuating phase is not chaotic
when parallel dynamics is used: specific configurations and not just
the magnetization converge to 2-cycles\footnote{There can be finitely many sites that oscillate with a larger period}.
(See also \citep{greil_period2_threshold_2007}.) The behavior of
$\langle\Omega\rangle$ and $\langle\Omega^{2}\rangle$ is not influenced
by these 2-cycles: all the contributions to moments come from the
fixed values of $\psi$ and $(\psi_{11},\psi_{10},\psi_{01},\psi_{00})$
even though when $C>C_{\mathrm{crit}}$ these are unstable fixed points
and the dynamics does not visit them.

Finally, we note that $\left\langle \Omega\right\rangle $ has interesting
behavior when $C\to\infty$ (after the limit $N\to\infty$): in graphs
without self-loops, $\left\langle \Omega\right\rangle $ diverges
at large $C$, while if self-loops are allowed, $\left\langle \Omega\right\rangle \to0$
at large $C$. This qualitative effect on the behavior caused by removing
self-loops is shown in Appendix \ref{sec:InhibitoryAppendix}.

\section{Discussion, open questions and future directions}

In this paper, we looked at the number and organization of fixed points
in different sparsely-connected cellular automata, and how their properties
are related to the dynamics. We have seen that at dynamical transitions
the fixed points divide up into multiple clusters upon crossing the
transition to the chaotic phase. The analysis shows a close connection
between low moments and the dynamics; $\langle\Omega\rangle$ is related
to the dynamics of the magnetization (see Eq. (\ref{eq:saddlepoint-1}))
and $\langle\Omega^{2}\rangle$ is related to the dynamics of the
distance between two copies.

The examples we studied illustrate different universality classes
of transitions. At each transition there is a certain change in the
dynamics which leads to a certain type of singularity in $\langle\Omega\rangle$
or $\langle\Omega^{2}\rangle$ as a function of $C-C_{c}$. The power
laws can be understood by considering how the form of the graph $q(\psi)$
or $\vec{q}(\vec{\phi})$ changes at the transition point (see Figs.
\ref{fig:excitatory_model} and \ref{fig:double-excitatory}). For
example, for the double excitatory model, Sec. \ref{subsec:The-Double-excitatory},
$\left\langle \Omega\right\rangle \sim\left(C-C_{c}\right)^{-\gamma}$
at $C\to C_{c}^{+}$ with $\gamma=1/2$. That the exponent $\gamma$
is equal to 1 on both sides of the transition in the excitatory model,
the inhibitory model and the Kauffman model can be justified by a
similar analysis. (This can also be used to calculate the ratio between
the coefficients on the two sides of the transition.) These predictions
apply to many systems and are independent of details such as the functions
used, the vertex degrees, and whether there are loops or not.

In the frozen phase, the structure of the graphs and in particular
the short loops, allows us to reproduce the moments from a different
perspective, and give even more information. In particular, they give
the entire distribution $P(\Omega)$, (see Eq. (\ref{eq:probability distribution})
and Fig. \ref{fig:P_Omega_frozen}) and show that for the Kauffman
model, the probability of having a fixed point $P(\Omega>0)\sim\sqrt{C_{c}-C}$.
In the fluctuating phase, we can determine only moments of the number
of fixed points, so we do not know how $P(\Omega>0)$ changes as $C\rightarrow C_{c}^{+}$.

The work raises several interesting directions for future study. In
the fluctuating phase, less is understood compared to the frozen phase,
since we do not understand the structure of the fixed points in terms
of changes in small subgraphs made of cycles and trees. We found that
there are multiple clusters of similar FPs, such that FPs belonging
to two different clusters involve extensive changes. What are the
statistics of the number of clusters or of the number of fixed points
in a cluster? Are there cases where the number of clusters diverges
at the transition as well as the number of fixed points belonging
to each cluster? Also, what is the probability of having no FPs near
the transition (in analogy with the frozen phase)? We know only the
moments of $\Omega$ in this phase, so we do not know whether $P(\Omega>0)\sim\sqrt{\left|C_{c}-C\right|}$
also in this phase. Another interesting thing to consider is how the
distributions of $\Omega$ differ between frozen and fluctuating phases.
We have found that in many cases the moments of $\Omega$ behave in
a similar way in both phases, although the algorithm for eliminating
frozen sites leaves behind an extensive number of sites with overlapping
loops. Also the dynamics is very different. In particular, in the
fluctuating phases, contributions to the moments often come from unstable
fixed points of $q$ or $\vec{q}$. Is the full distribution function,
unlike the low moments, very different in the fluctuating phase?

Another direction regards higher moments. For example, are there models
where the transition and associated singularities in the number of
FPs are only found in $\left\langle \Omega^{3}\right\rangle $ or
higher moments?

Here we have discussed properties only of FPs. Of course, there remain
very interesting but potentially very hard problems beyond FPs, such
as the distribution of period lengths and basin sizes.
\begin{acknowledgments}
We would like to thank Eric Brunet and Zhan Shi for discusssions on
the properties of the Lambert function and Paul D. Hanna for advice
about the sum appearing in the inhibitory model.
\end{acknowledgments}

\appendix

\section{Saddle points and mean field dynamics\label{sec:Saddle-points_from_MF_dynamics}}

The expressions for the expected number of fixed points can be interpreted
in a conceptual way that helps understand their behavior at transitions
for different models. We consider Boolean network models, with input
sites chosen at random, such as the models discussed in this article.

For such models, there is a relationship between the function $q(\psi)$
that describes the time dependence of the magnetization, $\psi=\frac{N_{1}}{N}$,
and the mean number of fixed points, see Eq. (\ref{eq:firstmoment-qsum}).
The formulas for the models studied here, the Kauffman model, double
excitatory model, single excitatory model, and inhibitory model have
this form, with different $q$'s. Notice that this formula is for
the case where the edges coming into each site are chosen completely
randomly and independently of one another; loops and double edges
are not ruled out. Allowing loops changes the number of fixed points
by a factor of order 1 (see the calculation for the inhibitory model
for an example). The multiple edges can be shown to have no effect
when $N$ is large.

As explained in the main text, the sum is dominated by values in the
vicinity of solutions to $\psi=q(\psi).$ Except when $\psi\approx0$
or $\psi\approx1$, the sum can be replaced by an integral. The Gaussian
approximation to the binomial distribution gives the contribution
near one of the fixed points $\psi_{i}$: 
\begin{equation}
\int_{\psi_{i}-\epsilon N}^{\psi_{i}+\epsilon N}dN_{1}\frac{1}{\sqrt{2\pi Nq(1-q)}}e^{-\frac{(N_{1}-q\left(\frac{N_{1}}{N}\right)N)^{2}}{2q(1-q)N}},\label{eq:movinggaussian}
\end{equation}
where all factors of $q$ stand for $q(\psi_{i})$. (Except for the
factor of $q(N_{1}/N)$ in the numerator, evaluating $q$ at the saddle
point $\psi_{i}$ leads to a vanishing error.) Defining $x$ by $N_{1}=N\psi_{i}+x$,
we find that $\left(N_{1}-q(\frac{N_{1}}{N})N\right)^{2}\approx x^{2}(1-q'(\psi_{i}))^{2}$.
This makes the width of the Gaussian different by a factor of $\frac{1}{|1-q'(\psi_{i})|}$
without changing the height of it, so the integral is 
\begin{equation}
\langle\Omega\rangle_{\psi_{i}}=\frac{1}{|1-q'(\psi_{i})|}.\label{eq:gaussian1d}
\end{equation}

The contributions to $\langle\Omega^{k}\rangle$ that can be described
by integrals can be calculated the same way. This is the same as the
\emph{first} moment of the number of fixed points in $k$ identical
copies of the system. One can think of the replicas as being one system
where each site has $2^{k}$ values. So to be more general, consider
a system with $l$ values and determine the first moment of $\langle\Omega\rangle$
for it; applying this to $k$ replicas of a system gives the moments
of the number of fixed points for it.

For a system with $l$ values define a magnetization vector $\vec{\psi}$
with $l$ components giving the fraction of sites with each of the
$l$ values. The dynamics of $\vec{\psi}$ is determined by a vector
of $l$ functions, $\vec{\psi}_{t+1}=\vec{q}(\vec{\psi}_{t})$. The
expected number of fixed points is given by a sum generalizing Eq.\ref{eq:firstmoment-qsum},
and the Gaussian approximation shows that near any $\vec{\psi}_{i}$
that is a fixed point of $\vec{q}$, whose components are nonzero,
the expected number of fixed points \citep{Turner_FPs_2025} is 
\begin{equation}
\left.\left(\left|\mathrm{det}\ \mathds{1}-\frac{(\partial\vec{q})'}{(\partial\vec{\psi})'}\right|\right)^{-1}\right|_{\vec{\psi}_{i}}\label{eq:fixed-points-lstates}
\end{equation}
where $\frac{(\partial\vec{q})'}{(\partial\vec{\psi})'}$ is the Jacobian
matrix of $l-1$ components of $\vec{q}(\vec{\psi})$ as a function
of the corresponding $l-1$ components of $\vec{\psi};$ since the
variables $\vec{\psi}$ sum to 1, only $l-1$ of them are independent
variables. (One can show that any $l-1$ components can be used).
To evaluate a contribution to $\langle\Omega^{2}\rangle$ this formula
can be applied to the functions $q_{11},q_{10},q_{01},q_{00}$ of
$\psi_{11},\psi_{10},\psi_{01},\psi_{00}.$

The Gaussian approximation to $\langle\Omega\rangle$ does not apply
if $\psi_{i}=0$ or 1 and the approximation to $\langle\Omega^{2}\rangle$
does not apply if one of the components of $(\psi_{11}^{(i)},\psi_{10}^{(i)},\psi_{01}^{(i)},\psi_{00}^{(i)})$
is zero, but the moments can still be written in terms of just the
derivatives of $q$ or $\vec{q}$ at the extremum. For $\langle\Omega\rangle$,
if the summand in Eq. (\ref{eq:firstmoment-qsum}) is written as $e^{NA(\psi)}$
and there is an extremum at an edge of the range of values for $A$,
then the first derivative does not have to be zero. E.g., near $\psi=0$
say $A(0)=0,A'(0)<0$, then $e^{NA(\psi)}\approx e^{-N|A'(0)|\psi}\approx e^{-|A'(0)|N_{1}}$.
This decays exponentially over a scale of order 1, so the sum cannot
be approximated by an integral.

If $\psi_{i}=0$ satisfies $q(\psi)=\psi$ then for small $N_{1}$'s:
\[
\binom{N}{N_{1}}q\left(\frac{N_{1}}{N}\right)^{n}\left(1-q(\frac{N_{1}}{N})\right)^{N-N_{1}}\approx\frac{N^{N_{1}}}{N_{1}!}\left(q'(0)\frac{N_{1}}{N}\right)^{n}\left(1-q'(0)\frac{N_{1}}{N}\right)^{N-N_{1}}\approx\frac{\left[q'(0)e^{-q'(0)}\right]^{N_{1}}N_{1}^{N_{1}}}{N_{1}!},
\]
which is more complicated than the exponential $e^{-|A'(0)|N_{1}}$
predicted at first (because $A$ is not an analytic function near
$N_{1}=0$) but is still independent of $N$ and decays over a scale
of order 1 as $N_{1}\rightarrow\infty$. So by Eq. (\ref{eq:Lambertderivative}),
\begin{equation}
\langle\Omega\rangle_{\psi\approx0}\approx\sum_{N_{1}}\frac{e^{-q'(0)N_{1}}q'(0)^{N_{1}}N_{1}^{N_{1}}}{N_{1}!}=\frac{1}{1+W\left(-q'(0)e^{-q'(0)}\right)}.\label{eq:LambertFrozen}
\end{equation}
Notice that when $q'(0)<0$ this is equal to $\frac{1}{1-q'(0)}$
by Eq. (\ref{eq:sinarcsin})), which agrees with the result of the
Gaussian approximation, although the sum is not Gaussian!

Similarly even near extrema where at least one of the components of
$(\psi_{11}^{(i)},\psi_{10}^{(i)},\psi_{01}^{(i)},\psi_{00}^{(i)})$
is zero, there are expressions for $\langle\Omega^{2}\rangle$ that
depend only on the derivatives of $\vec{q}$ at the extremum. The
sums cannot be summed immediately in terms of the Lambert function
(they are sums over multiple variables). See Appendix \ref{sec:generalsum}
for the expressions.

\section{Facts regarding the Lambert function\label{sec:Lambert_W}}

The Lambert W function\citep{Corless1996}, denoted by $W(x)$, is
used several times throughout the text. It is defined as the solution
to
\begin{equation}
W(x)e^{W(x)}=x.\label{eq:lambert}
\end{equation}
 For some $x$'s, this equation has two solutions; $W(x)$ is defined
as the solution which is zero at $x=0$ and whose value is in the
range $-1\leq W<\infty$, sometimes called $W_{0}$. This function
is real for $x\geq-1/e.$

Here are several useful relations that involve $W(x)$:

\begin{equation}
W(ye^{y})=y\mathrm{\ if}\ y\geq-1\label{eq:sinarcsin}
\end{equation}
 The series expansion near 0 is

\[
W(x)=x-x^{2}+\frac{3x^{3}}{2}-\frac{8x^{4}}{3}+\cdots=\sum_{n\ge1}(-)^{n-1}x^{n}\frac{n^{n-1}}{n!},
\]
see \citep{Melzak1973}, Ch. 3, Sec. 10, pg. 114. This formula is
connected to Cayley's formula for counting trees with $n$ numbered
vertices (see \citep{Polya1987} pg. 48 and \citep{Lovasz2007} chapter
4). The derivative satisfies:

\begin{equation}
\frac{1}{1+W(x)}+x\frac{dW(x)}{dx}=1\label{eq:derivativeofW}
\end{equation}
as can be seen by differentiating Eq. (\ref{eq:lambert}). Substituting
the series for $W(x)$ gives:
\begin{equation}
\frac{1}{1+W(x)}=\sum_{n\ge0}(-)^{n}x^{n}\frac{n^{n}}{n!}.\label{eq:Lambertderivative}
\end{equation}

\section{Sum Expressions for $\langle\Omega^{2}\rangle_{\mathrm{near}}$\label{sec:generalsum}}

Consider the number of pairs of fixed points with a fraction of sites
that are assigned state 1 equal to $\psi^{(i)}$. This corresponds
to an extremum $(\psi_{11},\psi_{10},\psi_{01},\psi_{00})=(\psi^{(i)},0,0,1-\psi^{(i)})$
in the sum for $\langle\Omega^{2}\rangle$, with components that are
zero, so the Gaussian approximation cannot be used. An expression
that gives the number of pairs of FPs in terms of derivatives of $\vec{q}$
nearby will now be derived. Such sums can also be derived for saddle
points $\vec{\psi}^{(i)}$where other combinations of components of
$\vec{\psi}$ are 0. These saddle points occur in families of automata
where there are FPs with two different fractions $\psi^{(1)},\psi^{(2)}$.
For example, if $\psi^{(1)}=0$, then the correlation between the
number of fixed points at $\psi\approx0$ and at $\psi\approx\psi^{(2)}$
is described by a saddle point at $\vec{\psi}=(0,0,\psi^{(2)},1-\psi^{(2)})$.
(This occurs in the excitatory model has this form, but it is easier
to determine $\langle\Omega^{2}\rangle$ from the distribution of
$\Omega$.)

We must evaluate the sum (\ref{eq:secondmoment-sum}) near $(\psi^{(i)},0,0,1-\psi^{(i)})$.
Since $N_{11}+N_{10}+N_{01}+N_{00}=1$, let us take $N_{11},N_{10},N_{01}$
as independent variables. The sum will be over finite values of $N_{10},N_{01}$
and $N_{11}/N$ close to $\psi^{(i)}$. Let us defne $k=N_{10},l=N_{01},n=N_{11},\bar{\theta}=n/N$,
and 
\[
t_{n,k,l}=\binom{N}{k,l,n,N-k-l-n}q_{10}^{k}q_{01}^{l}q_{11}^{n}q_{00}^{N-n-k-l}.
\]
Consider the effects of $k,l$ on the values of $q_{10}=q_{10}(\frac{n}{N},\frac{k}{N},\frac{l}{N}),$
etc. If two configurations are the same aside from a small number
of sites then applying the automaton to them will give two configurations
that are also close to each other, so $q_{10}(\frac{n}{N},0,0)=0$.
Also $q_{11}(\frac{n}{N},0,0)=q(\frac{n}{N})$ because when the configurations
are identical, they evolve by the dynamics of a single configuration.
The result of expanding $q_{10}(\frac{n}{N},\frac{k}{N},\frac{l}{N}$)
to first order in$\frac{k}{N},\frac{l}{N}$ is
\begin{align}
q_{10} & =\frac{1}{N}a(k,l,\bar{\theta}):=\frac{k}{N}\frac{\partial q_{10}}{\partial\psi_{10}}(\bar{\theta},0,0)+\frac{l}{N}\frac{\partial q_{10}}{\partial\psi_{01}}(\bar{\theta},0,0)\nonumber \\
q_{01} & =\frac{1}{N}b(k,l,\bar{\theta}):=\frac{k}{N}\frac{\partial q_{01}}{\partial\psi_{10}}(\bar{\theta},0,0)+\frac{l}{N}\frac{\partial q_{01}}{\partial\psi_{01}}(\bar{\theta},0,0)\label{eq:expansion2}\\
q_{11} & =q(\bar{\theta})+\frac{1}{N}c(k,l,\bar{\theta})\nonumber \\
q_{00} & =1-q(\bar{\theta})+\frac{1}{N}d(k,l,\bar{\theta})\nonumber 
\end{align}
where $c,d$ are not written out since they will not matter, aside
from satisfying the condition $a+b+c+d=0$ because the total sum of
$q_{ij}$ is 1.

The term $t_{n,k,l}$ can be simplified by factoring it: $\binom{N}{n,k,l,N-n-k-l}\left(\frac{a}{N}\right)^{k}\left(\frac{b}{N}\right)^{l}q_{11}^{n}q_{00}^{N-n-k-l}=\binom{N}{k,l,N-k-l}\left(\frac{a}{N}\right)^{k}\left(\frac{b}{N}\right)^{l}\times\binom{N-k-l}{n,N-n-k-l}q_{11}^{n}q_{00}^{N-n-k-l}$.
The first factor is approximately $\frac{a^{k}b^{l}}{k!l!}$. To approximate
the second factor assume that $n=Nq_{11}+x\sqrt{N}$ so that $N-n-k-l=N(1-q_{11})-x\sqrt{N}=Nq_{00}+a+b-k-l-x\sqrt{N},$and
then use Stirling's formula to approximate the binomial coefficient.
This gives $\frac{1}{\sqrt{2\pi Nq_{11}(1-q_{11})}}e^{-a-b-\frac{x^{2}}{2q_{11}(1-q_{11})}}$.
This is a Gaussian distribution, aside from $e^{-a-b}$. This combines
with the first factor to give $\frac{a^{k}b^{l}}{k!l!}e^{-a-b}$ ,
showing that $k$ and $l$ have a Poisson distribution\footnote{To give this an intuitive interpretation, consider $\binom{N}{n',k',l',N-n'-k'-l'}q_{10}(n,k,l)^{k'}q_{01}(n,k,l)^{l'}q_{11}(n,k,l)^{n'}q_{00}(n,k,l)^{N-n'-k'-l''},$which
is the probability that if a configuration has $n,k,l,N-n-k-l$ sites
assigned 11,10,01,00 respectively, that after one step of evolution
it will have $n',k',l',N-n'-k'-l'$ sites with these values. The second
moment is the sum over this after letting $n=n',k=k',l=l'$. If $k$
and $l$ are small, $k'$ and $l'$ have a Poisson distribution because
each site has only a small chance of being connected to a site in
the state 10 or 01 and being caused to change to 10 or 01, and these
sites are independent of one another.}. So
\[
t_{n,k,l}\approx\left[\frac{e^{-a-b}a^{k}b^{l}}{k!l!}\right]\left[\frac{1}{\sqrt{2\pi Nq_{11}(1-q_{11})}}e^{-\frac{N(\bar{\theta}-q(\bar{\theta}))^{2}}{2q_{11}(1-q_{11})}}\right],
\]
where the approximation $x=\frac{1}{\sqrt{N}}(n-Nq_{11}(n/N,k/N,l/N))\approx\frac{1}{\sqrt{N}}(n-Nq(n/N))$
has been made. This result can be derived conceptually by saying that
the probability that specific $k$ sites have the value 10 and specific
$l$ sites have the value $01$ is $\left(\frac{a}{N}\right)^{k}\left(\frac{b}{N}\right)^{l}$,
the chance that no other sites have these values is $(1-\frac{a+b}{N})^{N-k-l}\approx e^{-a-b}$,
the number of possibilities for these $k+l$ sites is $\binom{N}{k,l,N-k-l}\approx\frac{N^{k+l}}{k!l!}$.
These factors multiply to give the Poisson distribution. The $e^{-a-b}$
comes from $q_{11}^{n}q_{00}^{N-n-k-l}$ because this describes the
probability that certain sites are 11 or 00, Among the remaining sites
the number of sites assigned 11, $N_{11}$ has a Gaussian distribution
with a mean of $q_{11}(N-k-l)$, but the shift by $k+l$ can be neglected.

When this is summed over $k,l,n$, the values of $n$ where the second
factor is large are near solutions to $q(\bar{\theta})=\bar{\theta}$,
e.g., near $\psi^{(i)}$. The first factor is smooth as a function
of $\bar{\theta}$ since $k$ and $l$ are of order 1, so $\bar{\theta}$
can be replaced by $\psi^{(i)}$ in $a$ and $b$. Eq. (\ref{eq:expansion2})
implies $a=\kappa_{1}k+\kappa_{2}l$, $b=\kappa_{2}k+\kappa_{1}l$,
where $\kappa_{1}=\frac{\partial q_{10}}{\partial\psi_{10}}(\psi^{(i)},0,0),\kappa_{2}=\frac{\partial q_{10}}{\partial\psi_{01}}(\psi^{(i)},0,0)$.
So 
\begin{align}
\langle\Omega^{2}\rangle_{\mathrm{near,}\psi^{(i)}} & =\langle\Omega\rangle\sum_{k=0}^{\infty}\sum_{l=0}^{\infty}\frac{(k\kappa_{1}+l\kappa_{2})^{k}}{k!}\frac{(k\kappa_{2}+l\kappa_{1})^{l}}{l!}e^{-(\kappa_{1}+\kappa_{2})(k+l)},\nonumber \\
 & =\frac{1}{|1-q'(\psi^{(i)})|}\sum_{k=0}^{\infty}\sum_{l=0}^{\infty}\frac{(k\kappa_{1}+l\kappa_{2})^{k}}{k!}\frac{(k\kappa_{2}+l\kappa_{1})^{l}}{l!}e^{-(\kappa_{1}+\kappa_{2})(k+l)}\label{eq:directline}
\end{align}
where we have noticed that the sum over $n$ gives $\langle\Omega\rangle$,
as in Eq. (\ref{eq:movinggaussian}), which involves only the derivatives
of the $q$ functions at $(\psi^{(i)},0,0,1-\psi^{(i)})$.

\section{Determined Sites in the Kauffman Model\label{sec:frozen-sites}}

Sec. \ref{subsec:short-cycleFPS} described a method of searching
for fixed points of a Kauffman network. The algorithm finds certain
sites whose values are determined at a fixed point. If there are only
a small number of sites left over after this algorithm, then it allows
one to find all the fixed points (not missing fixed points with a
small basin of attraction for example) by searching through the combinations
of possible values for the remaining sites. This works in the frozen
phase of the Kauffman model. We here give an argument for why the
fraction of sites remaining after the action of the algorithm is zero
when $N\gg1$.

To see why this happens, we will give an argument similar to \citep{flyvbjerg1988},
see also \citep{Kaufman2005,mihaljev2006}. The algorithm for finding
which points are determined can be thought of as a system with three
values, $0,1,u$ (standing for ``unknown''). We will not erase the
arrows or the sites as in the version of the algorithm described above.
We start initially with all sites in the state $u$. Then when rule
(1)\emph{ }determines the value of a site, $u$ is replaced by this
value. This happens when each input site is definite or does not affect
the function for the site. Once a site is assigned 0 or 1, its value
will never change.

If the number of unknown inputs to a site is $k$, the chance that
it will be known at the next step is $2/\left(2^{2^{k}}\right)$;
$2^{2^{k}}$is the number of functions of the unknown variables, and
$2$ is the number of these that do not depend on the input. If $\psi_{u}$
is the fraction of unknown sites at a step, then the fraction of unknown
sites at the next step is:

\[
q_{u}(\psi_{u})=\sum_{k=0}^{\infty}\left(1-\frac{2}{2^{2^{k}}}\right)e^{-C\psi_{u}}\frac{(C\psi_{u})^{k}}{k!}.
\]
 This always has a fixed point at $\psi_{u}=0,$and it is attractive
when $C<2$ because $q_{u}'(0)=C/2$. This function does not seem
to have any other fixed points when $C<2,$ so as the process is iterated
$\psi_{u}\rightarrow0$, thus only a sub-extensive number of sites
is undetermined.

This algorithm gives a way to find the fixed points of the Kauffman
model in the frozen phase, but it seems likely that there can be frozen
phases of other models whose fixed points cannot be found this way.
For example, in a model where none of the input functions are constant
functions, there are no sites whose values are known for sure at the
beginning, yet some of these models seem likely to allow for frozen
phases. A simple example where no site's value is immediately determined,
but there is still a unique fixed point, is a sequence of $2n$ sites
with values $0,1$ where the functions are $x_{i}(t+1)=x_{i-1}(t)$
for $2\leq i\leq2n$ and $x_{1}(t+1)=x_{n}(t)+x_{2n}(t)-2x_{n}(t)x_{2n}(t)$.
(Every site's value depends on other sites' values, so there is no
site that is immediately determined. But one can see that nevertheless,
there is only one fixed point, where all sites are 0.)

A random network that seems also to have the behavior where the values
of sites at fixed points are determined although there is no way to
determine these values by backtracking for a small number of steps,
is one where all sites have two inputs and the functions are either
$x_{i}(t+1)=x_{j_{1}}(t)x_{j_{2}}(t)$ (with probability $p$) or
$x_{i}(t+1)=1-x_{j_{1}}(t)x_{j_{2}}(t)$ (with probability $1-p$)
. For $1>p>p_{c}$, one finds that there is only one saddle point
in the sum representing $\langle\Omega^{2}\rangle$ and it has $\psi_{01}=\psi_{10}=0$,
so that there is only one cluster of fixed points.

\section{Derivation details for the excitatory model\label{sec:excitatory_derivations}}

The excitatory model's behavior is determined entirely by topological
properties of random directed graphs, which have a transition similar
to the Erd\H{o}s\textendash Rényi transition in undirected graphs
where a large connected component forms at a critical connectivity,
see \citep{AlonSpencer2000} Chapter 10. (In this section we will
assume the random graphs are directed Erd\H{o}s\textendash Rényi graphs,
with loops allowed, rather than the graphs with multiple edges used
elsewhere in the paper. As stated above, this does not affect the
moments of $\langle\Omega\rangle$ when $N$ is large. This is checked
in footnote \ref{fn:doubleedge}.) The number of fixed points is the
number of ways to assign 0 and 1's to the sites of the directed graph
so that if $s$ is the value at a site and $s_{1},s_{2},\dots,s_{p}$
are the values at sites directed toward it, then $s=\max\{s_{1},s_{2},\dots,s_{p}\}$\textendash it
is the maximum of all its inputs. We note that loops provide the flexibility
that allows for multiple fixed points: On a tree, all sites must have
value 0, since, arguing by contradiction, if a site has value 1 there
must be an upstream path with all values 1, but any path in a tree
must end at a site with no inputs. On the other hand, a loop in isolation
can be assigned either all zeros or all ones, and the same applies
for an isolated SCC.

Consider a general directed graph. Not all sites belong to SCCs, since
a vertex might not belong to any directed loop. One can see that the
values for sites that do not belong to SCCs are determined by the
SCCs that are upstream from them\textendash they must be assigned
zero if all SCCs with paths leading to them have the value 0 (or if
there are no SCCs leading to them) and otherwise they are assigned
the value 1. Thus, values need only be chosen for the SCCs.

Now the values assigned to the SCCs have conditions on them. Consider
the reduced graph where each SCC is represented by a vertex, and there
is a directed edge between two such vertices if there is a directed
path connecting them. This graph does not have directed cycles in
it since all the SCCs on a cycle would in fact be a single SCC. On
this graph, if $s$ is the value assigned to an SCC and $s_{1},s_{2},\dots,s_{p}$
are the values of the SCCs directed toward it, then $s\geq\max\{s_{1},s_{2},\dots,s_{p}\}$.
Note that this is different from the constraint on the original graph
because of the inequality sign: an SCC can have the value 1 even if
all inputs are 0 because the cycles within itself allow this. The
number of fixed points is the number of ways to assign the values
to the SCCs, following this constraint.

The graph obtained by reducing a graph to the graph of its SCCs may
have a complex structure, but for an Erd\H{o}s\textendash Rényi graph,
it has a simple structure with probability one, and this makes it
easy to determine all the solutions to the above constraint. We first
note the difference between directed and undirected Erd\H{o}s\textendash Rényi
graphs. For undirected random graphs, when the number of edges is
greater than the critical number, there is a giant component which
contains a finite fraction of all vertices, while the other vertices
belong to components whose sizes are all small. In the directed graph,
there are different degrees of connectivity. Thus, besides the giant
SCC and vertices that are not connected to it by directed paths, there
are two other sets of vertices: vertices which have paths \emph{leading
to} the giant SCC and vertices with paths arriving at them \emph{from}
it. Each of these sets of sites is extensive. But the vertices that
are important for counting fixed points are the ones belonging to
SCCs, and besides the giant SCC, there are a finite number of these,
each containing one loop. They may be upstream, downstream, or disconnected
from the giant SCC, but there are no directed paths directly between
the small SCCs. The reason is that the chance of two small subgraphs
being connected by a \emph{short} path is $\sim1/N$ since the chance
of connections between specific vertices is $\sim1/N$. On the other
hand, if they are connected by a \emph{long} path it is likely that
some of the vertices on it belong to the giant SCC. Thus the structure
of the graph of SCCs is as indicated in Fig. \ref{fig:excitatory_model}(D).

This allows us to count the number of fixed points in the case $C<1$
and $C>1$, where $1$ is the point where the giant SCC appears. If
$C<1$, there are $O(1)$ SCCs, each with one cycle in it and no directed
paths between them, so each one can be assigned 0 or 1 independently.
Thus if there are $k$ SCCs, there are $2^{k}$ fixed points. When
there is a giant component, each cycle is constrained by the cycles
that are upstream from it. If there are $k_{up}$, $k_{down}$, and
$k_{disc}$ SCCs that are upstream, downstream and disconnected from
the giant SCC, then the SCCs can be counted as follows. If the value
of the giant component is 1, then the downstream components must have
the value 1, while if the giant component has value 0, the components
upstream from it must have the value zero. The rest of the components
can have arbitrary values. Thus the number of fixed points in the
two cases is $2^{k_{up}+k_{disc}}$ and $2^{k_{down}+k_{disc}}$respectively
or $2^{k_{disc}}(2^{k_{up}}+2^{k_{down}})$ fixed points altogether.

To understand the size of the giant component, consider any site of
the network. If one follows edges out from this, they form a tree
locally. The number of outgoing edges from each vertex in this tree
has a Poisson distribution with mean $C$. Thus, when $C<1$, the
mean number of descendants decreases exponentially with the generation,
so the tree ends. There can be no giant component then. If $C>1,$
then if the tree does not die out after a small number of steps, it
will keep growing and then it must contain points in the SCC. Let
$Q$ be the chance that a tree starting from a given site dies out
so that the site does not connect to the giant component. This may
be related to itself: for the vertex to not have a path toward the
giant component means that each site adjacent to it does not have
paths to the giant component. That is: 
\[
Q=\sum_{j=0}^{\infty}P(j\mathrm{\ descendants})Q^{j}=\sum_{j=0}^{\infty}\frac{e^{-C}(CQ)^{j}}{j!}=e^{C(Q-1)}
\]
The solution is $Q=-W(-Ce^{-c})/C.$ The probability to \emph{belong}
to the giant component is the probability that both the outgoing and
incoming trees from the site do percolate, $(1-Q)^{2}.$ Thus the
giant SCC has $(1-Q)^{2}N$ sites on average. Next, consider the number
of sites that have the same value as the giant component at the fixed
points, i.e. the sites downstream from it. These have connections
to the giant component in one direction but not the other, so there
are $Q(1-Q)N$ of them.

Next apply this reasoning to find the number of short cycles, and
hence the number of fixed points:

\textbf{Below the percolation transition, $C<1$}: The number of fixed
points is given by $\Omega=2^{k}$ where $k$ is the number of SCCs.
The number of cycles of length $n$ is Poisson distributed with mean
$\lambda_{n}=C^{n}/n$. Therefore, the mean of the total number of
cycles is
\[
\lambda=\sum_{n=1}^{\infty}\lambda_{n}=\ln\frac{1}{1-C},
\]
and the distribution of the total number is a Poisson distribution
also. So the probability distribution for the number of fixed points
is: 
\[
P\left(n\text{ FP}\right)=\begin{cases}
\left[1-C\right]\frac{\left(\ln\left(\frac{1}{1-C}\right)\right)^{k}}{k!} & n=2^{k}\\
0 & else
\end{cases}
\]

The moments are:
\begin{equation}
\left\langle \Omega^{r}\right\rangle =\sum_{k=0}^{\infty}2^{kr}P(2^{k}\text{FP})=\left(\frac{1}{1-C}\right)^{2^{r}-1}.\label{eq:excitatorybelowtranstion}
\end{equation}
This diverges at $C=1$ with exponent $2^{r}-1$.

\textbf{Above the percolation transition, $C>1$}: The number of fixed
points depends on $k_{\mathrm{up}},k_{\mathrm{down}},$ and $k_{\mathrm{discon}}$
as discussed above. To understand their statistics, note that a short
cycle will be in an SCC upstream from the giant component if at least
one of its sites has a macroscopic tree leaving it, and none have
a macroscopic tree going into it. (In the latter case it would be
part of the giant component). The sites that are upstream and downstream
from the SCC are not correlated, so this probability is $Q^{n}(1-Q^{n})$
where $n$ is the number of sites in the cycle. As different cycles
are uncorrelated, $k_{\mathrm{up}}$is a Poisson variable with the
mean $\lambda_{\mathrm{up}}=\sum_{n=1}^{\infty}\frac{C^{n}}{n}Q^{n}(1-Q^{n})=\log\frac{1-CQ^{2}}{1-CQ}$;
similarly for $k_{down}$. A cycle is unconnected to the giant component
with probability $Q^{2j}$, so $k_{disc}$ is a Poisson variable whose
mean is $\lambda_{\mathrm{discon}}=\mathrm{log}\frac{1}{1-CQ^{2}}$.

Thus, 
\begin{equation}
\langle\Omega\rangle=\langle2^{r_{\mathrm{\mathrm{discon}}}}(2^{r_{\mathrm{up}}+r_{\mathrm{down}}})\rangle=2e^{\lambda_{up}+\lambda_{discon}}=\frac{2}{1-CQ}=\frac{2}{1+W(-Ce^{-C})}.\label{eq:excitatoryabovetransition}
\end{equation}

The fixed points just found belong to two clusters: all the FPs where
the giant component has the value 1 are a finite distance from one
another, since only the values on short cycles upstream from the SCC
vary; likewise all the FPs where the giant component has the value
0 are a finite distance from one another. For the latter case, $\psi=0$
, while for the former, $\psi=(1-Q)^{2}+Q(1-Q)=1-Q$ since both the
sites in the giant component and downstream from it have the value
1; these agree with the two fixed points of $\psi\rightarrow q(\psi).$

Near and above the transition, where $C=1+\varepsilon$, one can show
that $Q=1-2\varepsilon+O\left(\varepsilon^{2}\right)$. Then $\left\langle \Omega\right\rangle \sim\frac{2}{C-1}$.
Below the transition the divergence is $\left\langle \Omega\right\rangle \sim\frac{1}{1-C}$.

Now we will calculate the expected number of fixed points using Eq.
(\ref{eq:firstmoment-qsum}), to compare with the above results. Recall
that we are not allowing double edges in this section (as mentioned
in Sec. \ref{subsec:Kauffman-model_moments} this does not affect
the answer). That is, each directed edge (including loops) is added
to the graph with a probability of $\frac{C}{N}$. We have
\begin{equation}
\langle\Omega\rangle=\sum_{N_{1}=0}^{N}{N \choose N_{1}}\left[\left(1-\frac{C}{N}\right)^{N_{1}}\right]^{N-N_{1}}\left[1-\left(1-\frac{C}{N}\right)^{N_{1}}\right]^{N_{1}}.\label{eq:excitatorysum}
\end{equation}
which has the form of Eq. (\ref{eq:firstmoment-qsum}) with $q(\psi)\approx1-e^{-C\psi}$
where $\psi=\frac{N_{1}}{N}$. Note that $q(\psi)=1-e^{-C\psi}$ is
the form that $q$ takes when multiple edges are allowed; thus multiple
edges do not affect $\langle\Omega\rangle$\footnote{Replacing $q=1-(1-\frac{C}{N})^{N_{1}}$ by $1-e^{-C\psi}$at first
seems to produce an error of order 1: the error $(1-\frac{C}{N})^{N_{1}}-e^{-C\psi}$
is of order $\frac{1}{N}$ (if $N_{1}$ is extensive). Since $q$
is raised to a power of order $N$ in Eq. (\ref{eq:excitatorysum}),
it seems that this discrepancy will change the result by a factor
of order 1. But the sum can be evaluated using Eq. \ref{eq:gaussian1d},
which does not involve raising $q$ to a large power, so the error
in the sum is negligible. The way both the exact sum and the approximate
sum come out the same is that their terms, graphed as a function of
$\psi$, are shifted by a small amount from one another. Since the
peak is narrow, this makes an order 1 difference to the terms, but
the total sums are equal. \label{fn:doubleedge}} . The maxima in this sum are the solutions to $\psi=q(\psi)$. One
solution is always $\psi_{1}^{*}=0$ while if $C>1$ there is a second
solution. If $C>1$, the former corresponds to the FPs where the giant
component and the sites upstream from it are assigned 0, while the
latter corresponds to the FPs where the giant component and the sites
downstream from it are assigned 1. The contribution to $\langle\Omega\rangle$
from $\psi\approx0$ is
\begin{equation}
\langle\Omega\rangle_{\psi\approx0}\approx\sum_{N_{1}=0}^{\infty}\frac{N^{N_{1}}}{N_{1}!}e^{-CN_{1}}\left(\frac{N_{1}C}{N}\right)^{N_{1}}\approx\sum_{N_{1}=0}^{\infty}\left(Ce^{-c}\right)^{N_{1}}\frac{N_{1}^{N_{1}}}{N_{1}!}\label{eq:lambertInhibitory}
\end{equation}
By equation (\ref{eq:Lambertderivative}) this is equal to $\frac{1}{1+W(-Ce^{-C})}$.
This gives 
\begin{align*}
\langle\Omega\rangle_{\psi\approx0} & =\begin{cases}
\frac{1}{1-C} & \mathrm{if}\ C<1\\
\frac{1}{1+W(-Ce^{-c})} & \mathrm{if}\ C>1,
\end{cases}
\end{align*}
which used Eq. (\ref{eq:sinarcsin}) when $C<1$. This agrees with
Eq. \ref{eq:excitatorybelowtranstion} when $C<1$ and half of Eq.
\ref{eq:excitatoryabovetransition} when $C>1$. .

The other half of the fixed points in Eq. (6) should come from the
other extremum. This extremum is the nonzero solution to $\psi=q(\psi),$
$\psi_{2}^{*}=1+W(-Ce^{-C})/C$. Note that this agrees with the magnetization
$1-Q$ of FPs with the giant component having the value 1. The number
of fixed points is expected to equal $\langle\Omega\rangle_{\psi\approx0}$
based on the symmetry between upstream and downstream components,
although the symmetry is not obvious from Eq. (\ref{eq:excitatorysum}):
the sum near $\psi_{2}^{*}=1-Q$ has a Gaussian form in contrast to
the sum giving $\langle\Omega\rangle_{\psi\approx0}$. However, using
Eq.(\ref{eq:gaussian1d}) shows that it is also equal to $\langle\Omega\rangle_{\psi\approx1-Q}=\frac{1}{1-Ce^{-C\psi}}=\frac{1}{1+W(-Ce^{-c})}$.

\section{Derivation details for the double-excitatory model\label{sec:Double-excitatory-derivations}}

\subsection{Dynamics}

Let $\psi_{n}$ be the mean magnetization (fraction of variables with
$x_{i}=1$) at time $n$. The dynamical update rule reads
\[
\psi_{t+1}=\sum_{k=0}^{\infty}P_{k}\left[1-\left(1-\psi_{t}\right)^{k}-k\psi_{t}\left(1-\psi_{t}\right)^{k-1}\right]\ ,
\]
where $P_{k}$ is the probability for $k$ incoming edges. For a Poisson
distribution, $P_{k}=e^{-C}C^{k}/k!$, so
\[
\psi_{t+1}=q(\psi_{t})=1-e^{-C\psi_{t}}\left(C\psi_{t}+1\right)
\]
This always has a fixed-point solution. For $C<C_{\mathrm{crit}}\simeq3.35$,
the $\psi=0$ solution is unique. For $C>C_{\mathrm{crit}}$, there
are three solutions and the middle one is unstable, so the dynamics
reach either the smallest or largest fixed ones, depending on the
initial conditions. See Fig. \ref{fig:double-excitatory}(A).

\subsection{Counting FPs}

The first moment, $\left\langle \Omega\right\rangle $, is given by
\begin{align*}
\left\langle \Omega\right\rangle  & =\sum_{N_{1}=0}^{\infty}\left(\begin{array}{c}
N\\
N_{1}
\end{array}\right)\left[\left(1-C/N\right)^{N_{1}}+N_{1}\frac{C}{N}\left(1-C/N\right)^{N_{1}-1}\right]^{N-N_{1}}\left[1-\left(1-C/N\right)^{N_{1}}-N_{1}\left(1-C/N\right)^{N_{1}-1}\frac{C}{N}\right]^{N_{1}}\\
 & \equiv\sum_{N_{1}=0}^{\infty}\Omega_{N_{1}}
\end{align*}
The first two terms are to be interpreted as $\Omega_{N^{*}=0}=1$
and $\Omega_{N^{*}=1}=0$. As in the last section, this formula assumes
multiple edges are not allowed, although this does not change the
results for $N>>1.$

We now consider the saddle point approximation to this. At $C<C_{\mathrm{crit}}$
the only contribution is at $\psi_{1}^{*}=N_{1}/N=0$. One can check
that only the $N_{1}=0$ term survives when $N\rightarrow\infty$,
for which $\Omega_{N_{1}=0}=1.$ (Even terms where $N_{1}$ remains
finite tend to zero.) At $C>C_{\mathrm{crit}}$, there are 3 contributions,
at the fixed points $\psi_{n+1}=\psi_{n}$ of the dynamical equation
above. The saddle-point around each of the two maxima at $\psi>0$
gives (by Eq. \ref{eq:gaussian1d})
\[
\frac{C\psi+1}{\left|(\psi-1)\psi C^{2}+\psi C+1\right|}
\]
so the result at $C>C_{\mathrm{crit}}$ is
\[
\left\langle \Omega\right\rangle =1+\sum_{i=2,3}\frac{C\psi_{i}^{*}+1}{\left|C^{2}\psi_{i}^{*}(\psi_{i}^{*}-1)+C\psi_{i}^{*}+1\right|}
\]
where $\psi_{2,3}^{*}$ are the two nonzero fixed points solutions.
This diverges at the transition, $C\to C_{\mathrm{crit}}^{+}$. Overall
there is a divergence only from one side of the transition, see Fig.
\ref{fig:double-excitatory}.. Also, the power law is $\langle\Omega\rangle\propto\frac{1}{(C-C_{\mathrm{crit}})^{1/2}}$
in contrast to the single excitatory model.

The $\langle\Omega\rangle\propto\frac{1}{(C-C_{\mathrm{crit}})^{1/2}}$
divergence can be derived from the general behavior of $q(\psi)$
near the transition. The graph of $q(\psi)$ first touches the line
$y=x$ at $C_{c}$. Say the point where they touch is $\psi_{0}$.
This implies that when $C=C_{c}+\epsilon,$ $q(\psi)-\psi_{0}=a(\epsilon)+b(\epsilon)(\psi-\psi_{0})-c(\psi-\psi_{0})^{2}$,
with $a(\epsilon)=a'_{0}\epsilon$ and $b(\epsilon)=1+b'_{0}\epsilon$.
The fixed points are at $\left(b'_{0}\epsilon\pm\sqrt{4a'_{0}c\epsilon}\right)/(2c)$
and the value of $\frac{dq}{d\psi}$ at them is $1\mp\sqrt{4a'_{0}c\epsilon}$,
so $\langle\Omega\rangle=\frac{2}{\sqrt{4a'_{0}c\epsilon}}.$

Without working out $\left\langle \Omega^{2}\right\rangle $, it must
have qualitatively the same behavior. For $C<C_{\mathrm{crit}}$,
there is exactly one FP, therefore $\left\langle \Omega^{2}\right\rangle =1$,
and above the transition $\left\langle \Omega^{2}\right\rangle \ge\left\langle \Omega\right\rangle ^{2}$
and so there will be a divergence. Whether $C$ is bigger or smaller
than $C_{\mathrm{crit}},$there is always a unique fixed point near
$\psi_{1}^{*}=0$, the one where all sites are off. As in Sec. \ref{subsec:short-cycleFPS},
any other fixed point nearby would differ from this at cycles with
trees going out from them. But then the sites of the graph do not
satisfy the condition of having inputs from at least two other sites
that are on, since all sites outside the subgraph are off. But there
can be pairs of FPs that are close to each other when $\psi>0$. We
have the following example: for a cycle with exactly one incoming
edge from an active site (from outside the cycle) to each vertex,
the entire cycle can be either on or off. As long as this does not
affect many sites downstream, this gives two nearby stationary configurations.

\section{Derivation details for the inhibitory model\label{sec:InhibitoryAppendix}}

\begin{figure}
\begin{centering}
\includegraphics[viewport=0bp 0bp 620bp 423bp,clip,width=0.5\textwidth]{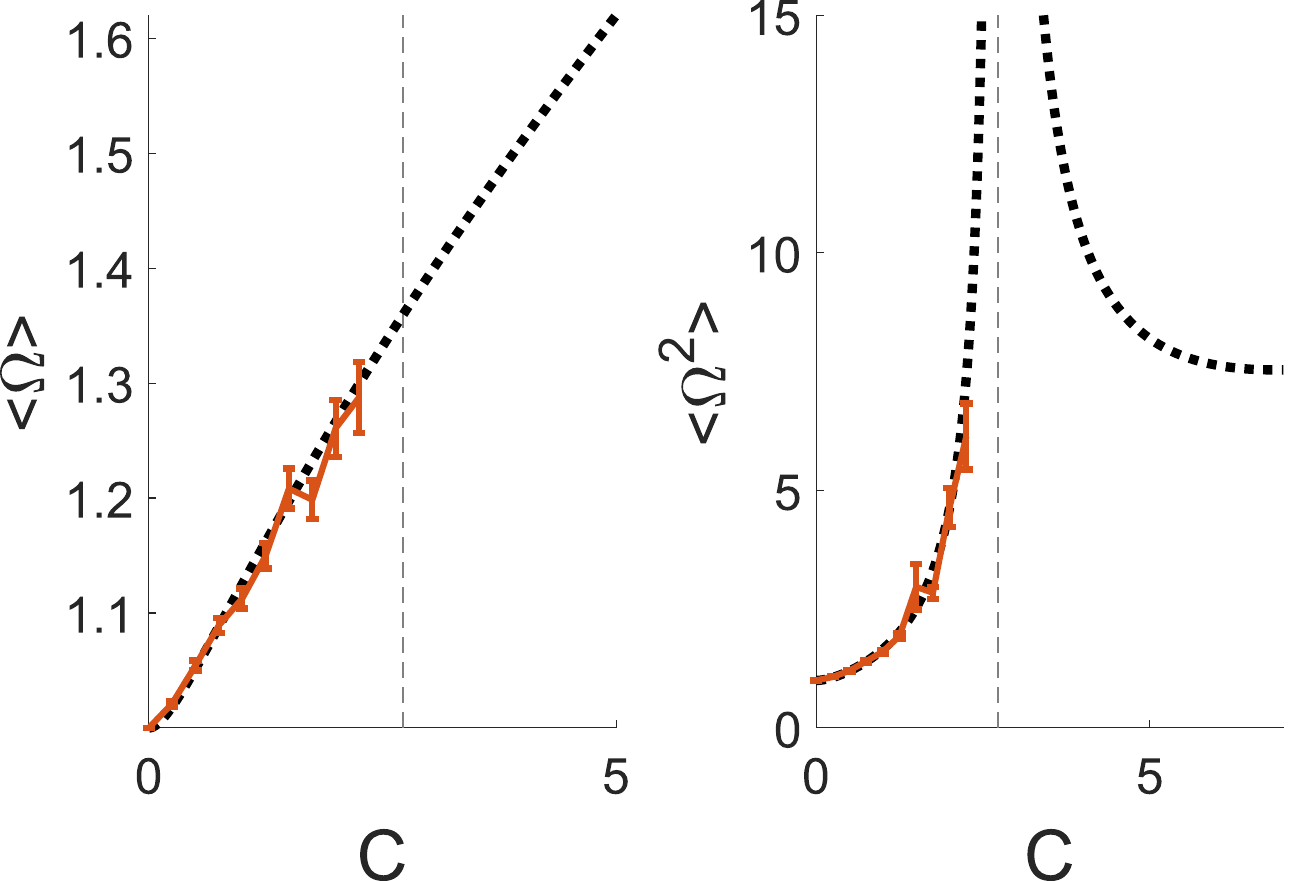}
\par\end{centering}
\caption{\label{fig:inhibitory_moments}Analytical results for $\left\langle \Omega\right\rangle $
and $\left\langle \Omega^{2}\right\rangle $ (dashed lines) for the
inhibitory model, compared to exhaustive numerical searches (error
bars) at $N=5000$.}
\end{figure}

In this Appendix we present the calculation of $\left\langle \Omega\right\rangle $
and $\left\langle \Omega^{2}\right\rangle $ for the inhibitory model,
as well as $P(\Omega)$ in the frozen phase. The results for $\left\langle \Omega\right\rangle $
and $\left\langle \Omega^{2}\right\rangle $ are plotted in Fig. \ref{fig:inhibitory_moments},
and compared to an exhaustive search algorithm in the frozen phase
($C<e$ in this model).

As for the Kauffman model, in the frozen phase we find that a graph
simplification algorithm allows to simplify large graphs with thousands
of sites, until exhaustive search for FPs is possible. It repeats
the following steps, that do not change the number of FPs, until no
further changes are possible:
\begin{enumerate}
\item Remove all sites $i$ that have no incoming edges, and all outgoing
edges from these sites (including the vertices that they point toward).
This can be done because a site $i$ with no inputs will have $x_{i}=1$
at all FPs, and if site $i$ is an input to site $j$ then at all
FPs $x_{j}=0$, so site $j$ does not affect any other site.
\item Remove all sites that have no outgoing edges.
\end{enumerate}

\subsection{1st moment}

Consider a configuration with $N_{1}$ sites that are on and $N-N_{1}$
that are off, and determine the probability that it will be a fixed
point when the connections are assigned. A site that is on will stay
on if all inputs are off. The probability is $\left(1-\frac{N_{1}}{N}\right)^{k}$
when there are $k$ inputs. Hence
\begin{equation}
q(\psi)=\sum_{k=0}^{\infty}e^{-C}\frac{C^{k}}{k!}(1-\frac{N_{1}}{N})^{k}=e^{-C\psi}.\label{eq:inhibitory_magnetisationdynamics}
\end{equation}
 (This assumes the graph is constructed as in Sec.\ref{subsec:Kauffman-model_moments},
with multiple edges allowed).

Thus the average number of fixed points is as in Eq. (\ref{eq:firstmoment-qsum}),
\begin{equation}
\langle\Omega\rangle=\sum_{N^{*}=0}^{N}\binom{N}{N_{1}}q\left(\frac{N_{1}}{N}\right)^{N_{1}}\left(1-q\left(\frac{N_{1}}{N}\right)\right)^{N-N_{1}}\equiv\sum_{N_{1}=0}^{N}\Omega_{N_{1}}.\label{eq:inhibitoryOmegasum}
\end{equation}
 The contribution to the saddle point approximation is at the solution
to $q(\psi)=\psi$ (see Sec. \ref{sec:Mean-magnetization}), namely
$\psi^{*}(C)=\frac{W\left(C\right)}{C}$ where $W$ is the Lambert
$W$ function, which gives (according to Eq. (\ref{eq:gaussian1d}
)
\[
\left\langle \Omega\right\rangle =\frac{1}{1+W\left(C\right)}
\]
This is an analytic function of $C$ with no singularities.

\begin{quotation}
\end{quotation}

\subsection{2nd moment\label{subsec:2nd-momentInhibitory}}

Take two configurations, with $N_{00},N_{10},N_{01},N_{11}$ the numbers
of sites with the pairs of values $00,10,01,11$. Also let $\psi_{00}=\frac{N_{00}}{N}$
etc. The probability that this is a fixed point is $q_{00}^{N_{00}}q_{01}^{N_{01}}q_{10}^{N_{10}}q_{11}^{N_{11}}$
where $q_{00}$ is the probability that the inputs to a site would
cause it to become 00 at the next step, etc. Then
\begin{equation}
q_{11}=\sum_{k=0}^{\infty}e^{-C}\frac{C^{k}}{k!}\psi_{00}^{k}=e^{-C(1-\psi_{00})}\label{eq:q11}
\end{equation}
where the sum is over the number of incoming edges. Also, 
\begin{equation}
q_{10}=\sum_{k}e^{-C}\frac{C^{k}}{k!}\left[(\psi_{01}+\psi_{00})^{k}-\psi_{00}^{k}\right]=e^{-C(1-\psi_{00})}\left[e^{C\psi_{01}}-1\right].\label{eq:q10}
\end{equation}
while the formula for $q_{01}$ is similar and $q_{00}=1-q_{11}-q_{01}-q_{10}.$
The 2nd moment is given by Eq. (\ref{eq:secondmoment-sum}). 
\begin{align}
\left\langle \Omega^{2}\right\rangle  & =\sum_{N_{11},N_{01},N_{10}}\left(\begin{array}{c}
N\\
N_{00},N_{01},N_{10},N_{01}
\end{array}\right)q_{11}^{N_{11}}q_{00}^{N_{00}}q_{01}^{N_{01}}q_{10}^{N_{10}}\equiv\sum_{N_{11},N_{01},N_{10}}\Omega_{N_{11},N_{01},N_{10}}\label{eq:sums-with-qs}
\end{align}
The extrema are at solutions to $\vec{\psi}=\vec{q}(\vec{\psi})$.
This condition always has a solution corresponding to two fixed points
in the same cluster, $\psi_{01}=\psi_{10}=0,\ \psi_{11}=\frac{W(C)}{C},$
and when $C>C_{\mathrm{crit}}=e$ another solution given by 
\begin{align}
\psi_{11} & =-\frac{W\left(-\frac{W^{2}\left(C\right)}{C}\right)}{C}\nonumber \\
\psi_{01} & =\psi_{10}=\frac{W\left(C\right)}{C}+\frac{W\left(-\frac{W^{2}\left(C\right)}{C}\right)}{C}.\label{eq:2nd-fixed-point}
\end{align}
(Notice that these equations are consistent with the fixed point that
occurs in one copy of the system: $\psi_{11}+\psi_{10}=\psi$ because
this counts the number of sites in the first copy with the value 1.)
This represents a pair of fixed points not in the same cluster. For
large $C$, $\psi_{11}\ll\psi_{01},\psi_{10}$, so there is little
overlap between FPs except for ones that are essentially identical.
In fact, $\psi$ (the fraction of sites equal to 1 in a typical fixed
point) is approximately $\frac{\log\!C}{C}$ and $\psi_{11}\approx(\frac{\log\!C}{C})^{2}$
so it is as if different fixed points are made up of independently
chosen sets.

Eq. \ref{eq:sums-with-qs} can be summed using the saddle point approximations
given above. The contribution from the first fixed point, $\psi_{10}=\psi_{01}=0,\psi_{11}=\frac{W(C)}{C}$
can be evaluated using Eq. \ref{eq:directline}, $\langle\Omega^{2}\rangle=\langle\Omega\rangle\sum_{k_{1},k_{2}}\frac{k_{1}^{k_{2}}k_{2}^{k_{1}}}{k_{1}!k_{2}!}\left(\frac{W(C)^{2}}{C}\right)^{k_{1}+k_{2}}$.
We see numerically that the series is equal to 
\begin{equation}
\sum_{k_{1},k_{2}}\frac{k_{1}^{k_{2}}k_{2}^{k_{1}}}{k_{1}!k_{2}!}y^{k_{1}+k_{2}}=\frac{1}{1-W^{2}\left(-y\right)}=\frac{1}{1-W^{2}\left(-\frac{W(C)^{2}}{C}\right)}\label{eq:cross-exponents}
\end{equation}
 This expression can be proved using interesting identities about
binomial coefficients (see Appendix \ref{sec:crossexponents}). Below
the transition, for $C<e$, this is the only saddle-point contribution
to the sum. Here, one can show\footnote{Using $W(C)e^{W(C)}=C$, it follows that $-\frac{W(C)^{2}}{C}=-W(C)e^{-W(C)}$.
So if $x=W(-\frac{W(C)^{2}}{C})$ then $xe^{x}=-W(C)e^{-W(C)}$. A
solution is $x=-W(C)$, which can be checked to be the correct branch
when $C<e$.} that $W\left(C\right)=-W\left(-\frac{W^{2}\left(C\right)}{C}\right)$,
so 
\[
\langle\Omega^{2}\rangle=\langle\Omega^{2}\rangle_{\mathrm{near}}=\langle\Omega\rangle\frac{1}{1-W^{2}(C)}=\frac{1}{\left(1+W(C)\right)^{2}\left(1-W(C)\right)}\mathrm{\ for\ }C<C_{\mathrm{crit}}.
\]

Above $C_{\mathrm{crit}}$, there is also the second contribution,
from near Eq. \ref{eq:2nd-fixed-point}, which can be evaluated using
Eq. \ref{eq:fixed-points-lstates}:
\[
\left(\langle\Omega^{2}\rangle_{\text{far}}\right)^{-1}=\left|\begin{array}{ccc}
1-\frac{\partial q_{11}}{\partial\psi_{11}} & \ \ -\frac{\partial q_{11}}{\partial\psi_{10}} & \ \ -\frac{\partial q_{11}}{\partial\psi_{01}}\\
-\frac{\partial q_{10}}{\partial\psi_{11}} & \ \ \ \ 1-\frac{\partial q_{10}}{\partial\psi_{10}} & \ \ -\frac{\partial q_{10}}{\partial\psi_{01}}\\
-\frac{\partial q_{01}}{\partial\psi_{11}} & \ \ -\frac{\partial q_{10}}{\partial\psi_{10}} & \ \ \ \ 1-\frac{\partial q_{01}}{\partial\psi_{01}}
\end{array}\right|=\left|\begin{array}{ccc}
1+Cq_{11} & \ \ Cq_{11} & \ \ Cq_{11}\\
Cq_{10} & \ \ \ \ 1+Cq_{10} & \ \ -Cq_{11}\\
Cq_{01} & \ \ -Cq_{11} & \ \ \ \ 1+Cq_{01}
\end{array}\right|=(1+C\psi)(1-C\psi_{11}),
\]
where first the derivatives of $q_{ij}$ are calculated and then the
conditions that $q_{11}=\psi_{11},q_{10}=\psi_{10}$, etc., at the
saddle point and $\psi=\psi_{10}+\psi_{11}$ are used to simplify
the expression. Substituting the values for $\psi$ and $\psi_{11}$
for the saddle point gives

\[
\langle\Omega^{2}\rangle=\langle\Omega^{2}\rangle_{\mathrm{far}}+\langle\Omega^{2}\rangle_{\mathrm{near}}=\frac{\langle\Omega\rangle^{2}}{\left(1+W\left(-\frac{W^{2}(C)}{C}\right)\right)}+\frac{\left\langle \Omega\right\rangle }{1-W^{2}\left(-\frac{W^{2}\left(C\right)}{C}\right)}\mathrm{\ for\ }C>C_{\mathrm{crit}}.
\]

These formulae give behavior at the transition that is very similar
to the Kauffman model: when $C\approx e$, the mean number of fixed
points is finite, $\langle\Omega\rangle\approx\frac{1}{2}$ and

\begin{align}
\langle\Omega^{2}\rangle & \approx\begin{cases}
\frac{e}{2(e-C)} & \mathrm{if}\ C<e\\
\frac{e}{C-e} & \mathrm{if}\ C>e
\end{cases}.\label{eq:inhibitory-singularity}
\end{align}

It is also interesting to consider the limit when $C$ is large: For
large $C$, $\langle\Omega\rangle=\frac{1}{1+W(C)}\approx\frac{1}{\ln C}$,
which tends to zero. The second moment is approximately the same as
the first moment, which can be understood if one assumes that it is
very unlikely for their to be more than one fixed point. We have also
calculated $\langle\Omega\rangle$ and $\langle\Omega^{2}\rangle$
when the model is modified so that loops are not allowed. The saddle
point calculations are very similar to the case where loops are allowed:
the saddle point has the same location, and there is a factor without
singularities multiplying the expression. For the inhibitory model,
we find that $\langle\Omega\rangle=\frac{C}{W(C)}\frac{1}{1+W\left(C\right)}$.
On the one hand, the extra factor of $\frac{C}{W(C)}$is not singular
at the transition point, so the qualitative behavior at the transition
is not affected by whether loops are allowed. On the other hand, the
limit at large $C$ is different, $\langle\Omega\rangle\approx\frac{C}{(\ln C)^{2}}$,
tending to infinity; when loops are allowed and $C$ is large, many
sites have edges to themselves, so they cannot be on at a FP, and
this constraint rules out most possibilities for FPs and makes $\langle\Omega\rangle$
tend to zero instead.

Below the transition one can calculate the full distribution of $\Omega$;
As for the Kauffman model, the probability of a fixed point existing
tends to zero at $C_{\mathrm{crit}}$.

\subsection{Dynamics}

Now we consider the dynamics briefly. The time dependence of the magnetization
changes its behavior at $C_{\mathrm{crit}}$. To see this, note that
a fixed point $\psi^{*}$ of $\psi\rightarrow q(\psi)$ is attractive
if $|q'(\psi^{*})|<1$. Calculating $q'(\psi^{*})$ from Eq. (\ref{eq:inhibitory_magnetisationdynamics})
and the value of $\psi^{*}$, one sees that the derive $q'(\psi^{*})$
is between $-1$ and 0 when $C<e$ and is less than -1 when $C>e$.
Hence the fixed point begins to repel $\psi$ when $C>e$. One can
check that there is a 2-cycle that $\psi$ is attracted to in this
case.

On the other hand, the mean number of fixed points $\langle\Omega\rangle=\frac{1}{|1-q'(\psi^{*})|}$
does not have any nonanalytic behavior at $e$. For one, $\psi^{*}(C)=W(C)/C$
varies analytically near $C=e$, and for two $q'(\psi^{*})=-1$, so
$\langle\Omega\rangle$ does not diverge.

Now consider the dynamics of $\text{\ensuremath{\vec{\psi}=}}(\psi_{11},\psi_{10},\psi_{01},\psi_{00})$.
We saw above that it has a unique fixed point, $(W(C)/C,0,0,1-W(C)/C)$,
when $C<e$. This fixed point attracts all initial values of $\vec{\psi}$.
This has a conequence for the dynamics of specific configurations:
it implies that they are all attracted to a common fixed point (up
to finitely many sites) since the distance between any two tends to
$\psi_{10}+\psi_{01}=0$. I.e., this is a frozen state. When $C>e$,
this fixed point becomes unstable and $\vec{\psi}$ is attracted to
2-cycles instead (which agrees with the instability in the dynamics
of the magnetization).

At the same $C$, there is also a transition in the FPs from one cluster
to multiple clusters. Unlike for the Kauffman model, the distance
between FPs in different clusters is not related to the value of $\psi_{10}+\psi_{01}$
for the steady-state value of $\vec{\psi}$ that the system is attracted
to. This does not even make sense, because $\vec{\psi}$ is not attracted
to a steady-state but to 2-cycles! The distance between FPs from different
clusters was found in Sec. \ref{subsec:2nd-momentInhibitory} to be
given by the value of $\psi_{01}+\psi_{10}$ at a second fixed point
that appears at $C=e$. (This point is also unstable, but it still
is important for the properties of the FPs.) These observations show
that the transition in the dynamics and the transition in the properties
of the fixed points could possibly happen at different $C_{\text{crit}}$'s
(if the model is changed slightly). One transition is related to the
appearance of the attractive 2-cycles and the other is related to
the appearance of a fixed point with $\psi_{01}+\psi_{10}\neq0$.

\section{The sum for the Inhibitory Model\label{sec:crossexponents}}

The sum $f(x)=\sum_{k_{1},k_{2}=0}^{\infty}\frac{k_{1}^{k_{2}}k_{2}^{k_{1}}}{k_{1}!k_{2}!}x^{k_{1}+k_{2}}=\frac{1}{1-W(-x)^{2}}$
has an interesting analytical derivation, which we found with the
help of an entry in the Online Encyclopedia of Integer Sequences \citep{HannaOeis}
and advice from its author, Paul Hanna. The entry is about the sequence
of numbers
\[
a_{n}=\sum_{j=0}^{n}\binom{n}{j}j^{n-j}(n-j)^{j}.
\]
This integer is essentially the sequence of coefficients of $f(x)$:
After substituting $n=k_{1}+k_{2}$and $j=k_{1}$ in the definition
of $f(x)$, $f(x)=\sum_{j=0}^{n}\frac{a_{n}x^{n}}{n!}$. The article
states the formula $f(x)=\frac{1}{1-W(-x)^{2}}$ (due to Paul Hanna).
The proof is not given but the formula can be derived from another
interesting formula for $a_{n}$ given in the article (due to Vladeta
Jovovic \citep{HannaOeis}) :
\begin{equation}
\frac{a_{n}}{n!}=\frac{n^{n}}{n!}-\frac{n^{n-1}}{(n-1)!}+\frac{n^{n-2}}{(n-2)!}-\dots+\frac{(-1)^{n}}{0!}.\label{eq:taylorseries}
\end{equation}

So first let us prove Eq. \ref{eq:taylorseries} and then derive the
formula for $f(x)$. We begin with

\begin{equation}
n!=\sum_{j=0}^{n}(-1)^{n-j}\binom{n}{j}(\sigma+j)^{n},\label{eq:differentiating}
\end{equation}
where $\sigma$ can be any number. This can be found in books about
the calculus of finite differences: Define the difference operator
on a function, $\Delta g(\sigma)=g(\sigma+1)-g(\sigma)$. By calculating
the differences of the differences iteratively, $\Delta^{k}\sigma^{n}$
is a polynomial whose degree is $n-k$ and whose leading coefficient
is $n(n-1)(n-2)\dots(n-k+1)$, for $k\leq n$. If $k=n$, this means
$\Delta^{n}\sigma^{n}$ is a constant and the constant is $n!$ The
operator $\Delta$ can be written as $D-I$ where $D$ is a shift
operator, $Dg(\nu)=g(\nu+1)$, so $\Delta^{n}=\sum_{j=0}^{n}(-1)^{n-j}\binom{n}{j}D^{j}$
by the binomial theorem. So $\Delta^{n}\sigma^{n}=n!$ is the same
as Eq. (\ref{eq:differentiating}).

Now substitute this formula into the definition of $a_{n}$:

\begin{align*}
a_{n} & =\sum_{j=0}^{n}\binom{n}{j}j^{n-j}\left[\sum_{k=0}^{j}\binom{j}{k}n^{k}(-j)^{j-k}\right]\\
 & =\sum_{0\leq k\leq j\leq n}(-1)^{j-k}\binom{n-k}{j-k}\binom{n}{n-k}j^{n-k}n^{k}\\
 & =\sum_{0\leq l\leq s\leq n}(-1)^{l}\binom{s}{l}\binom{n}{s}\left(l+n-s\right)^{s}n^{n-s}\\
 & =\sum_{s=0}^{n}(-1)^{s}\binom{n}{s}s!n^{n-s}.
\end{align*}
The third step changed variables to $s=n-k$ and $l=n-j$, and the
last step used Eq. (\ref{eq:differentiating}). This is the same as
Eq. (\ref{eq:taylorseries})

Now this can be substituted into $f(x)$ and summed using the Taylor
series for $W(x)$. 
\[
f(x)=\sum_{d=0}^{\infty}(-1)^{d}\sum_{n=d}^{\infty}\frac{n^{n-d}x^{n}}{(n-d)!}.
\]
 Now $W(x)^{d}=\sum_{n=d}^{\infty}(-1)^{n-d}d\frac{n^{n-d-1}x^{n}}{(n-d)!}$
(see Ref. \citep{Corless1996}, Eq. 2.36) and differentiating with
respect to $x$ gives $\frac{W(x)^{d}}{(1+W(x))}=\sum_{n=d}^{\infty}(-1)^{n-d}\frac{n^{n-d}x^{n}}{(n-d)!}$
(using Eq. (\ref{eq:derivativeofW})). Hence 
\[
f(x)=\sum_{d=0}^{\infty}\frac{W(-x)^{d}}{1+W(-x)}=\frac{1}{1-W(-x)^{2}}.
\]

\bibliographystyle{unsrt}
\bibliography{Thesis}

\end{document}